\documentclass[apj,twocolumn]{emulateapj}

\pdfoutput=1

\usepackage{epsfig}
\usepackage{graphicx}
\usepackage{amsthm}
\usepackage{amssymb}
\usepackage{rotating}
\usepackage{multirow}
\usepackage{longtable}
\usepackage{savesym}
\savesymbol{tablenum}
\restoresymbol{SIX}{tablenum}
\usepackage{booktabs}

\defcitealias{2019ApJ...873...51L}{Paper~I}

\revised{\today}
\shorttitle{Surveying G\ion{H}{2} Regions: II. M\,17}
\shortauthors{Lim, De\,Buizer, \& Radomski}

\begin{document}
%\LongTables

\title{Surveying the Giant \ion{H}{2} Regions of the Milky Way with \textit{SOFIA}: II. M\,17}

%\correspondingauthor{Wanggi Lim}
\email{wlim@usra.edu}

\author{Wanggi Lim, James M. De Buizer, and James T. Radomski}
\affil{\footnotesize \textit{SOFIA}-USRA, NASA Ames Research Center, MS 232-12, Moffett Field, CA 94035, USA}

\begin{abstract}
We present our second set of results from our mid-infrared imaging survey of Milky Way Giant \ion{H}{2} regions. We used the FORCAST instrument on the Stratospheric Observatory For Infrared Astronomy to obtain 20 and 37\,$\mu$m images of the central $\sim$10$\arcmin\times$10$\arcmin$ area of M\,17. We investigate the small- and large-scale properties of M\,17 using our data in conjunction with previous multi-wavelength observations. The spectral energy distributions of individual compact sources were constructed with {\it Spitzer}-IRAC, \textit{SOFIA}-FORCAST, and {\it Herschel}-PACS photometry data and fitted with massive young stellar object (MYSO) models. Seven sources were found to match the criteria for being MYSO candidates, four of which are identified here for the first time, and the stellar mass of the most massive object, UC\,1, is determined to be 64\,$M_{\sun}$. We resolve the extended mid-infrared emission from the KW Object, and suggest that the angle of this extended emission is influenced by outflow. IRS\,5 is shown to decrease in brightness as a function of wavelength from the mid- to far-infrared, and has several other indicators that point to it being an intermediate mass Class II object and not a MYSO. We find that the large-scale appearance of emission in M\,17 at 20\,$\micron$ is significantly affected by contamination  from  the  [SIII]  emission line from the ionized gas of the giant \ion{H}{2} region. Finally, a number of potential evolutionary tracers yield a consistent picture suggesting that the southern bar of M\,17 is likely younger than the northern bar.

\end{abstract}


\section{Introduction} 

This is the second paper in a series of studies of the infrared properties of galactic giant \ion{H}{2} (G\ion{H}{2}) regions. An overview of the nature of G\ion{H}{2} regions and why they are important to study is highlighted in detail in the introduction of Lim \& De Buizer (2018; hereafter ``\citetalias{2019ApJ...873...51L}''). To summarize in brief, G\ion{H}{2} regions are areas within galaxies where the majority of high-mass star formation is occurring. Galaxies like the Milky Way contain on the order of 50 of these regions and the bolometric flux of entire host galaxy is dominated by their emission. They are identified by being extremely bright in the infrared due to the high levels of heating of their dusty environments, and by their bright cm radio continuum emission, due to the copious amount of Lyman continuum photons that the young OB stars produce ($10^{50}-10^{52}$ LyC photons s$^{-1}$;\citealt{2004MNRAS.355..899C}). The emitting region of the radio continuum from these sources is typically quite large, spanning a few to 10s of pc in size. G\ion{H}{2} regions are useful laboratories for the study of high-mass star formation as well as star cluster formation within starburst-like galactic environments. 

Our original source list comes from \citet{2004MNRAS.355..899C} who published an article identifying all 56 of the bona-fide G\ion{H}{2} regions in our Galaxy. We aim to compile a 20 and 37\,$\mu$m imaging survey of as many of these G\ion{H}{2} regions within the Milky Way as we can with the \textit{Stratospheric Observatory For Infrared Astronomy} (\textit{SOFIA}) and its mid-infrared instrument FORCAST, creating complete and unsaturated maps of these regions with the best resolution ever achievable at our longest wavelength (i.e., $\sim$3$\arcsec$ at 37\,$\mu$m). From our observations of these sources individually and as a group we will gain a better understanding of their physical properties individually and as a population. 

In this paper we will concentrate on the G\ion{H}{2} region M\,17. At a distance of 1.98\,kpc \citep{2011ApJ...733...25X}, M\,17 is the closest G\ion{H}{2} region to Earth and consequently has been the subject of numerous studies. At optical wavelengths the region is dominated by the reflection nebula known as the Omega Nebula (or Swan Nebula) due to its overall shape. Located within this optical nebulosity is a young \citep[$<$10$^6$\,yr;][]{1997ApJ...489..698H} open cluster called NGC\,6618, whose $\sim$100 O and B-type stars \citep[e.g.,][]{1980A&A....91..186C, 1997ApJ...489..698H, 2008ApJ...686..310H} are responsible for the heating and reflected emission seen as the Omega Nebula. The central stars in this cluster are also mostly responsible for the ionization of the G\ion{H}{2} region here, which physically separates the two major extended infrared structures known as the northern bar, or M\,17\,N (a.k.a. ``M\,17-North''), and southern bar, or M\,17\,S (a.k.a. ``M\,17-South''). M\,17\,S is bordered to the southwest by a large, dense molecular cloud referred to as M\,17\,SW \citep{1975ApJ...195..367L}. M\,17\,S is also the transition region between the \ion{H}{2} region and the M\,17\,SW molecular cloud, and is therefore perhaps the best-studied edge-on photodissociation region (PDR) in the Galaxy \citep[e.g.,][and references therein] {2007ApJ...658.1119P}. M\,17 contains three compact sources that have been extensively studied: the region's brighest radio peak, UC\,1 \citep{1984A&A...136...53F}; its radio-quiet but equally bright infrared neighbor IRS\,5 \citep{2000A&A...357L..33C}; and the so-called Kleinmann-Wright Object \citep[``KW'':][]{1973ApJ...185L.131K}, which is a very bright infrared source that is isolated and to the southwest of the larger infrared emitting areas of the M\,17 nebula. 

\begin{figure*}[htb!]
\epsscale{1.15}
\plotone{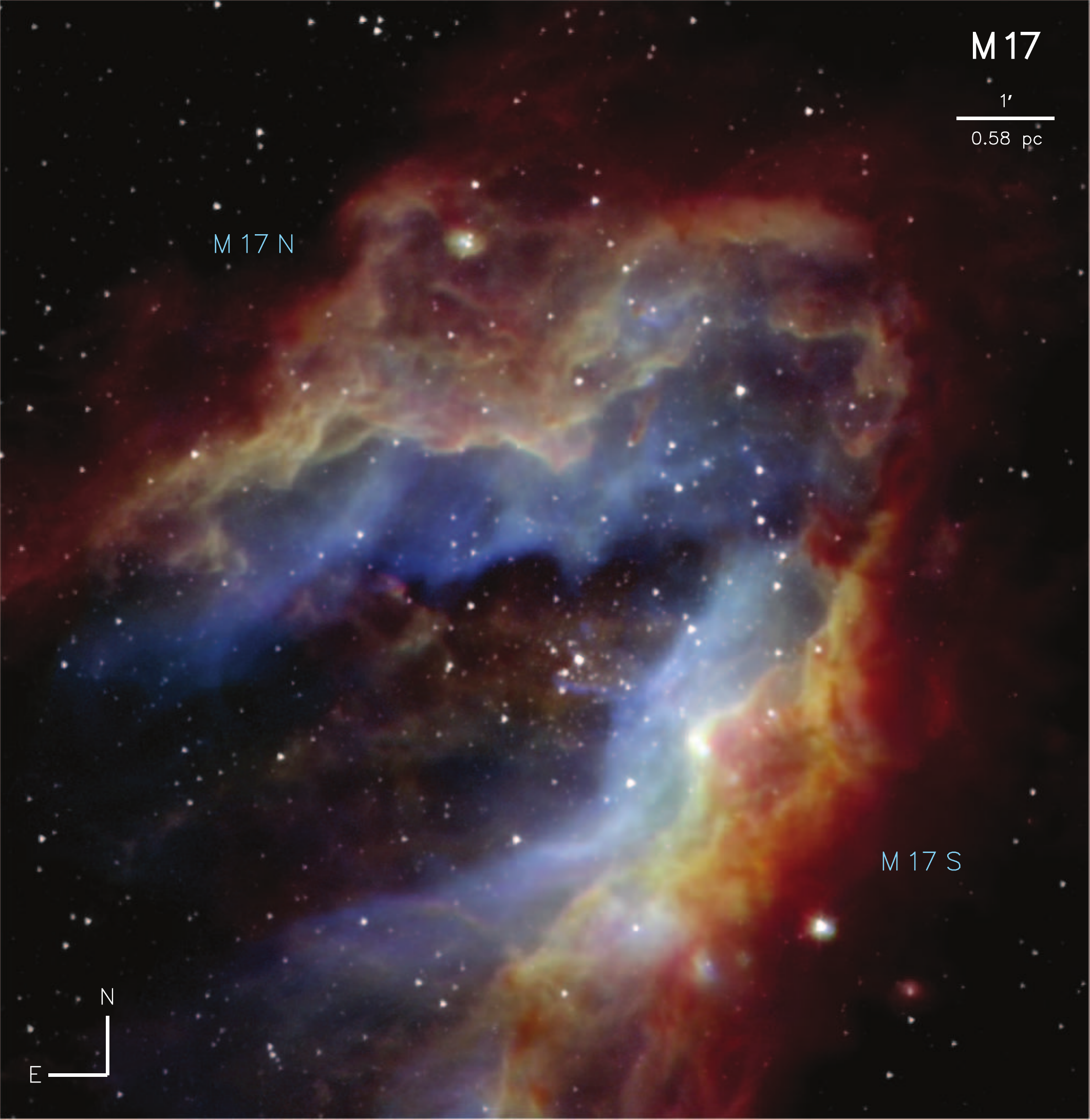}
\caption{A 3-color image of a $\sim10\arcmin\times10\arcmin$ field centered on M\,17. Blue is the \textit{SOFIA}-FORCAST 20\,$\mu$m image, green is the \textit{SOFIA}-FORCAST 37\,$\mu$m image, and red is the \textit{Herschel}-PACS 70\,$\mu$m image. Overlaid in white is the \textit{Spitzer}-IRAC 3.6\,$\mu$m image, which traces the revealed stars within M\,17, field stars, and hot dust}. \label{fig:m17}
\end{figure*}

Like our previous target of study (W\,51\,A; \citetalias{2019ApJ...873...51L}) M\,17 is sufficiently large, complicated, and well-studied that we devote to it this entire paper. In the next section (Section \ref{sec:obs}), we will discuss the new \textit{SOFIA} observations and give information on the data obtained for M\,17. In Section \ref{sec:results1}, we will give more background on this region as we compare our new data to previous observations and discuss individual sources and regions in-depth. In Section \ref{sec:data}, we will discuss our data analysis, modeling, and derivation of physical parameters of sources and regions. Our conclusions are summarized in Section \ref{sec:sum}.

\section{Observations and Data Reduction} \label{sec:obs}

The observational techniques and data reduction processes employed on the M\,17 data were, for the most part, identical to those described in \citetalias{2019ApJ...873...51L} for W\,51\,A. We will highlight below some of observation and reduction details specific to the M\,17 observations, however for a more in-depth discussion of these details and techniques, refer to \citetalias{2019ApJ...873...51L}. 

The data presented here for M\,17 was obtained during \textit{SOFIA}'s Cycle 5 using the FORCAST instrument \citep{2013PASP..125.1393H} on the nights of 2017 August 2 (Flight 425), 2017 September 26 (Flight 433), and 2017 September 27 (Flight 434). In an attempt to expand the spatial coverage of our infrared maps out even further from the center of the G\ion{H}{2} region, M\,17 was revisited in the summer of 2018 during Cycle 6. However, due to an instrumentation issue, the Cycle 6 data could not be properly corrected for distortions in the optical plane, making it impossible to incorporate into the Cycle 5 data map. This outer region of M\,17 was found to consist of extended dust emission structures, though there is one point-like source in the Cycle 6 data (which we name Source 1), which is likely associated with an ammonia clump named MSX6C G014.9790-00.6649 found by \citet{2011MNRAS.418.1689U}. Since the photometric calibration of the Cycle 6 data for Source 1 is good, we will include it in our analyses of compact objects within M\,17 in Section \ref{sec:data}. We also included a sub-frame of a Cycle 6 pointing towards the KW object, since it was partially off the edge of the array in the Cycle 5 data. 

FORCAST is a dual-array mid-infrared camera capable of taking simultaneous images at two wavelengths. The short wavelength camera (SWC) is a 256$\times$256 pixel Si:As array optimized for 5–-25\,$\mu$m observations; the long wavelength camera (LWC) is a 256$\times$256 pixel Si:Sb array optimized for 25–-40\,$\mu$m observations. After correction for focal plane distortion, FORCAST effectively samples at 0$\farcs$768 pixel$^{-1}$, which yields a 3$\farcm$4$\times$3$\farcm$2 instantaneous field of view. Observations were obtained in the 20\,$\mu$m ($\lambda_{eff}$=19.7\,$\mu$m; $\Delta\lambda$=5.5\,$\mu$m) and 37\,$\mu$m ($\lambda_{eff}$=37.1\,$\mu$m; $\Delta\lambda$=3.3\,$\mu$m) filters simultaneously using an internal dichroic. 

All images were obtained at aircraft altitudes between 39,000 and 41,000 feet and by employing the standard chop-nod observing technique used in the thermal infrared, with chop and nod throws sufficiently large to sample clear off-source sky (typically $\sim$7$\arcmin$). The mid-infrared emitting region of M\,17 G\ion{H}{2} is much larger ($\sim$9$\arcmin\times$9$\arcmin$) than the FORCAST field of view, and thus had to be mapped using multiple pointings. We created a mosaic from 11 individual pointings, with each pointing having an average on-source exposure time of about 180s at both 20\,$\mu$m and 37\,$\mu$m. Images from each individual pointing was stitched together using the \textit{SOFIA} Data Pipeline software REDUX \citep{2015ASPC..495..355C} as a test for producing FORCAST LEVEL 4 imaging mosaics.

Flux calibration for each of the 11 individual pointings was provided by the \textit{SOFIA} Data Cycle System (DCS) pipeline and the final total photometric errors in the mosaic were derived using the same process described in \citetalias{2019ApJ...873...51L}. The estimated total photometric errors are 15\% for 20\,$\mu$m and 10\% for 37\,$\mu$m. 
All images then had their astrometry absolutely calibrated using \textit{Spitzer} data by matching up the centroids of point sources in common between the \textit{Spitzer} and \textit{SOFIA} data. Absolute astrometry of the final \textit{SOFIA} images is assumed to be better than 1$\farcs$5, which is a slightly more conservative estimate than that quoted in \citetalias{2019ApJ...873...51L} (i.e. 1$\farcs$0) due to slight changes in the focal plane distortion and our ability to accurately correct it with the limited calibration data available for these observations.

In order to perform photometry on mid-infrared point sources, we employed the aperture photometry program \textit{aper.pro}, which is part of the IDL DAOPHOT package available in The IDL Astronomy User's Library (http://idlastro.gsfc.nasa.gov).

\begin{figure*}[htb!]
\epsscale{1.1}
\plotone{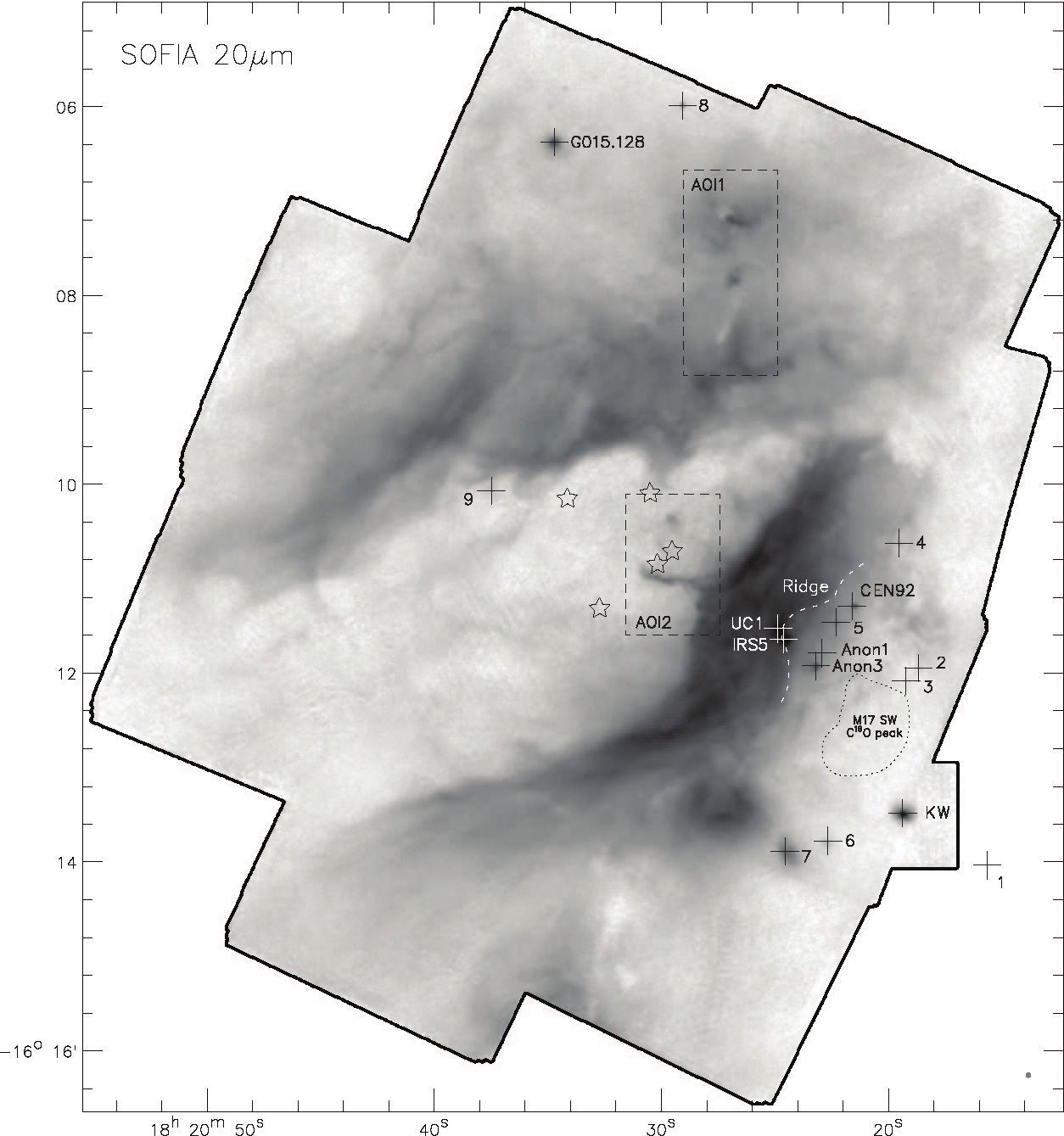}
\caption{M\,17 image mosaic taken at 20\,$\mu$m by \textit{SOFIA} shown in inverse color (i.e. brighter features are  darker in color). Sources discussed in the text are labeled. Areas of Interest (AOI1, AOI2) which are discussed in Appendix \ref{appendixb} and shown in Figure \ref{fig:aoiboth} are surrounded by dashed boxes. The highest contour of C$^{18}$O emission from \citep{2003ApJ...590..895W}, is shown as a reference for the location of the center of the M\,17\,SW molecular cloud (dotted line). The star symbols represent the 5 most massive ($>$O7) stars in the central 1$\arcmin$ radius of the open cluster NGC\,6618 (from \citealt{1997ApJ...489..698H}). The gray dot in the lower right indicates the resolution of the image at this wavelength.  \label{fig:all19}}
\end{figure*}

\begin{figure*}[htb!]
\epsscale{1.1}
\plotone{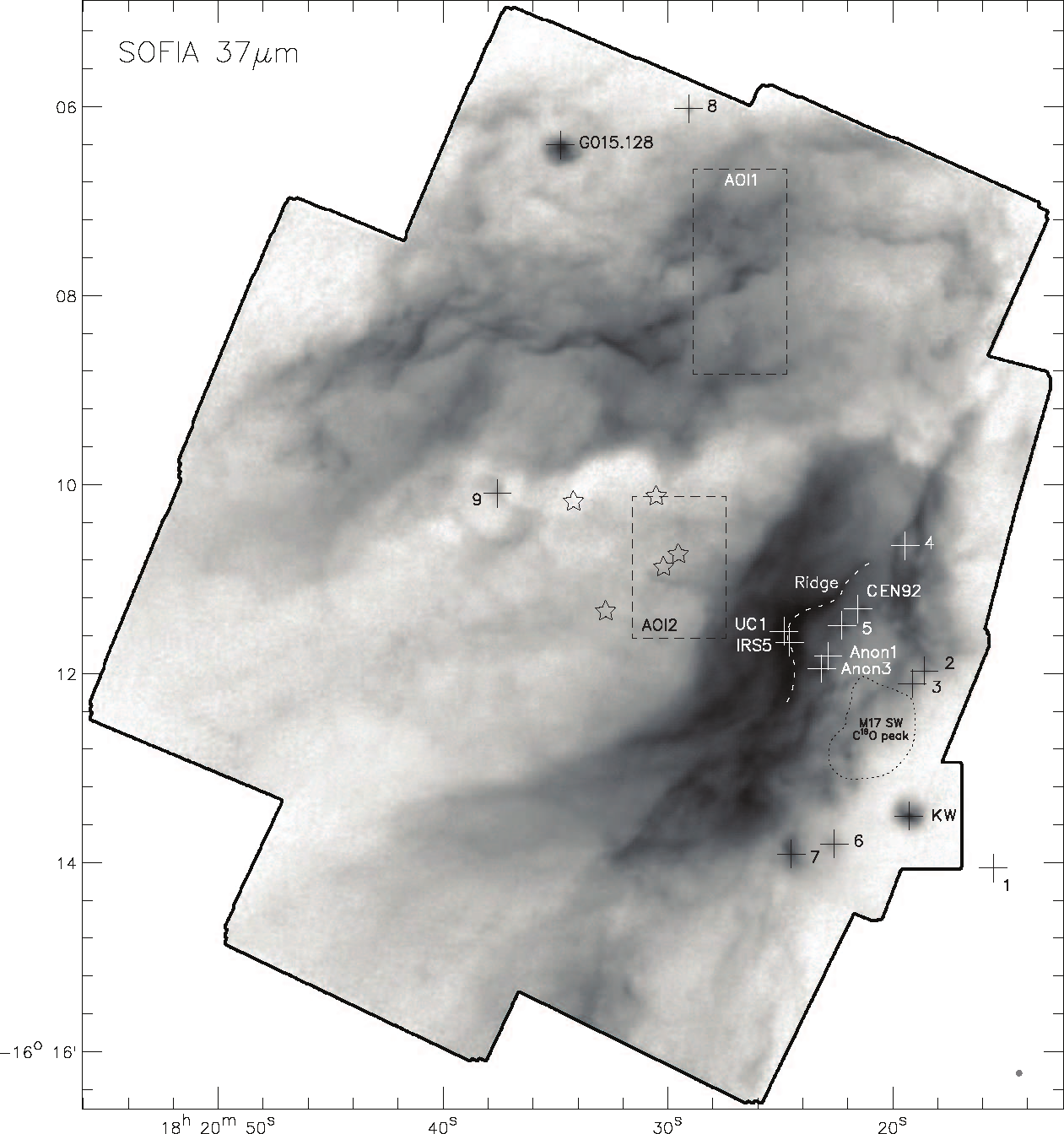}
\caption{M\,17 image mosaic taken at 37\,$\mu$m by \textit{SOFIA} shown in inverse color (i.e. brighter features are  darker in color). Labeling is the same as in Figure \ref{fig:all19}. \label{fig:all37}}
\end{figure*}

\section{Comparing \textit{SOFIA} Images to Previous Imaging Observations} \label{sec:results1}

The large-scale extended 20 and 37\,$\mu$m emission of M\,17 (Figure \ref{fig:m17}) covers predominantly the same area as the wide-spread 21\,cm continuum emission as seen by \citet{1984A&A...136...53F}, however there are major differences in the internal structure of the emission seen at the two infrared wavelengths. The 20\,$\mu$m image of M\,17 shows brighter emission towards center of the nebula, and grows fainter with radius (Figure \ref{fig:all19}). The 37\,$\mu$m emission brightens further from the center of the nebula (Figure \ref{fig:all37}). The naive assumption for this difference would be dust temperature, with the 20\,$\mu$m tracing the hotter dust closer to the ionizing stars of NGC\,6618 near the center of the nebula, and the 37\,$\mu$m tracing cooler dust at a larger distance. However, there is actually a much greater correlation in morphology and brightness distribution of the dust in the 37\,$\mu$m map with the \textit{Spitzer} images at 3.6--5.8\,$\mu$m and even the \textit{Herschel} 70\,$\mu$m image, than the 20\,$\mu$m map. This is most evident in Figure \ref{fig:m17}, where blue is represented by the 20\,$\mu$m data and is much more prominent in the inner regions of the nebula, whereas the 37\,$\mu$m (green) and 70\,$\mu$m (red) emission are much more co-spatial. We have discovered in the data from this survey that the fluxes measured in the 20\,$\mu$m filter of the FORCAST instrument on \textit{SOFIA} can often have enhanced emission due to the presence of a very strong [\ion{S}{3}] emission line at 18.71\,$\mu$m when looking at ionized regions. For M\,17,  this is evidenced by the spectra from \textit{ISO} taken at different locations from near the center of the nebula to the southwest across the M\,17\,S bar (see Appendix \ref{appendixd}). These spectra not only show the presence of bright [\ion{S}{3}] emission within the nebula,  but they also show a trend where the line strength grows as you approach the brightest areas of 20\,$\mu$m emission, reaching line fluxes of thousands of Jy above the dust continuum. Therefore, the main reason why the large-scale 20\,$\mu$m morphology looks significantly different from the images at all other infrared wavelengths in M\,17 is likely due to enhanced flux from the emission of [\ion{S}{3}] tracing the ionized gas being liberated from the inner walls of M\,17\,N and M\,17\,S that are facing the central O stars of the revealed NGC\,6618 stellar cluster. Another difference between the 20 and 37\,$\mu$m images is that the 20\,$\mu$m emission of M\,17\,S does not extend as much towards the southwest into the M\,17\,SW molecular cloud as the 37\,$\mu$m emission does, while the emission at 20 and 37\,$\mu$m is similar in overall extent for the M\,17\,N region. These effects are likely due to the fact that the overall extinction towards M\,17\,N is far less than M\,17\,S, as evidenced by the presence of optical and H$\alpha$ emission only associated with M\,17\,N \citep[e.g.,][]{1968PASJ...20...95I,1985MNRAS.216..761C}. Consistent with this are measurements towards M\,17 that show smaller visual extinctions varying across M\,17\,N with A$_V$ values between $\sim$0.4 and $\sim$6\,magnitudes, and larger A$_V$ values between $\sim$6.5 and $\sim$14.5\,magnitudes across M\,17\,S. \citep{2002ApJ...574..187A,2005ARep...49...36G}.

\subsection{Discussion of Individual Sources}\label{sec:sources}

Given the expansive nature of the cm radio continuum from the G\ion{H}{2} region environment, it is difficult to detect and/or isolate possible emission from individual sources within M\,17, except for the bright emission from UC\,1. \citet{2012ApJ...755..152R} identified a few dozen compact cm continuum sources at 3.5 and 6\,cm with the \textit{JVLA}, but in addition to UC\,1 and KW, they only detect emission from one infrared-bright source seen in our 20 and 37\,$\mu$m data (CEN\,92, a.k.a. B\,331).  

The vast majority of the previously identified young stars discovered via near-infrared imagery \citep[e.g.,][]{1976A&A....50...41B,1980A&A....91..186C,1997ApJ...489..698H,2002ApJ...577..245J} are not detected in the 20 and 37\,$\mu$m images we present here. In fact, only one of the six young stellar object (YSO) candidates identified in the near-infrared by \citet{1997ApJ...489..698H} has detectable mid-infrared emission (CEN\,92). There have been mid-infrared studies identifying high-mass Class I sources or massive young stellar objects \citep[MYSOs; e.g.,][]{2001A&A...377..273N, 2004aap...427..849C}, but we do not detect most of them in our 20 and 37\,$\mu$m maps. Some of these sources may simply be misclassified, like the silhouette disk source M\,17-SO1 \citep{2004aap...427..849C} and massive Class I candidate CEN\,34 \citep{2001A&A...377..273N} which have since been reclassified as a low-mass object \citep{2005Natur.434..995S} and a background post-AGB star \citep{2013A&A...557A..51C}, respectively. However, some of these sources are likely to be more evolved Class II or III sources, and/or low-mass such that the emission from their circumstellar environment is too faint to be seen against the infrared background of the extended M\,17 G\ion{H}{2} region. 

In the subsections that follow, we will discuss several compact and individual objects of interest that were detected in the \textit{SOFIA} maps, and comment on new insights that these data bring to bear on their nature. We will also discuss other areas of interest within M\,17 in Appendix \ref{appendixb}.

\begin{figure*}[htb!]
\epsscale{0.80}
\plotone{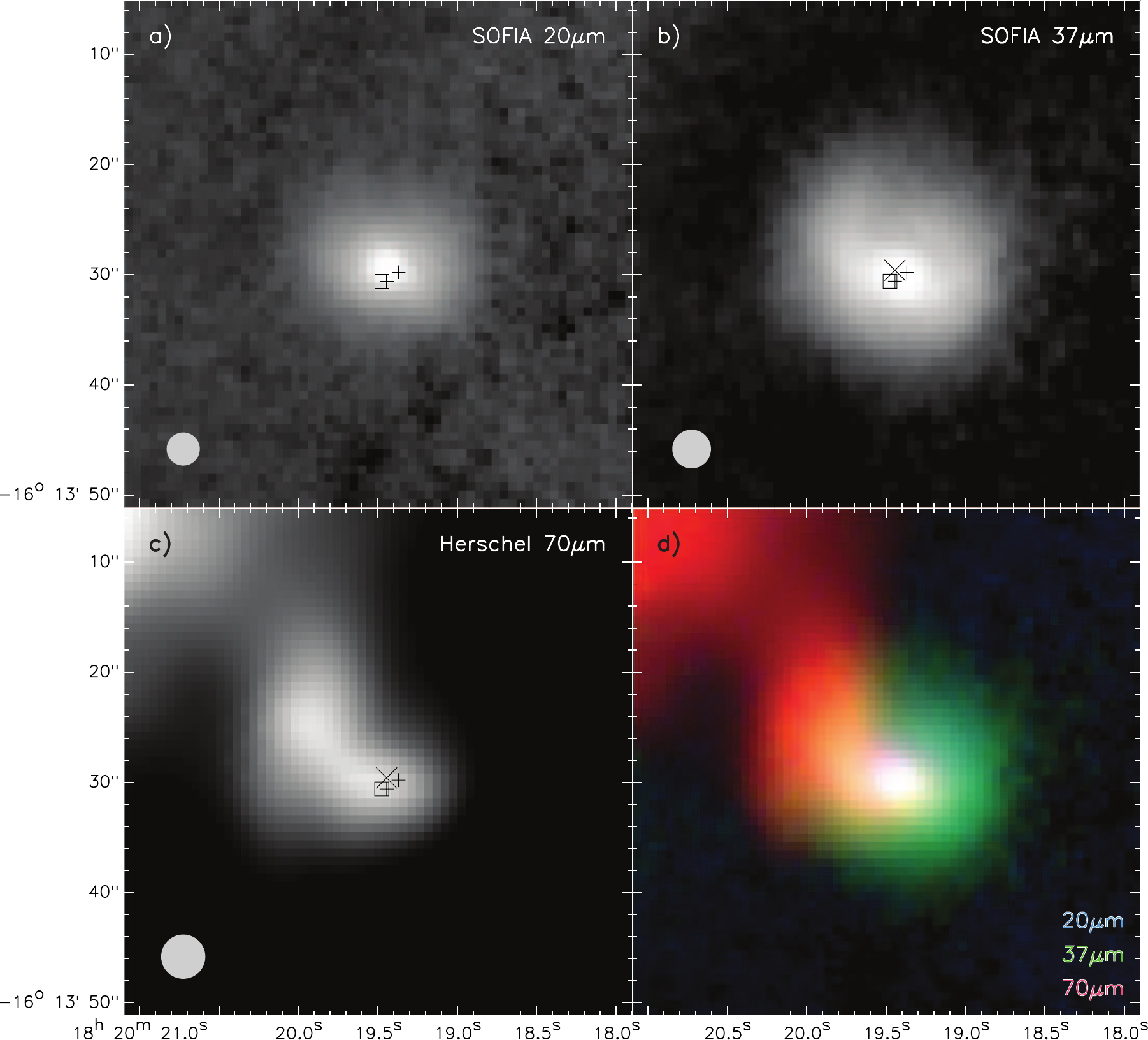}
\caption{The Kleinmann-Wright Object (a.k.a. M\,17SW\,IRS\,1).  a) \textit{SOFIA} 20\,$\mu$m image; b) \textit{SOFIA} 37\,$\mu$m image; and c) \textit{Herschel} 70\,$\mu$m image. The location of the 20\,$\mu$m peak is given by the X symbol, and the location of the radio continuum peak JVLA\,3 from \citet{2012ApJ...755..152R} is given by the square. The plus symbols show the positions of the two near-infrared sources from \citet{2004aap...427..849C}, with the northernmost source being their Source 2 (KW-2) and the southernmost Source 1 (KW-1). d) An RGB image with the wavelengths representing each color given in the lower right corner. The gray circles in the lower left of the panels show the spatial resolution of the images in those panels.\label{fig:kwo}}
\end{figure*}

\subsubsection{Kleinmann-Wright Object (a.k.a. M~17SW~IRS~1)}
The Kleinmann–Wright (KW) Object \citep{1973ApJ...185L.131K}, is a binary system seen in the near-infrared with a position angle of about 45$\arcdeg$ \citep{2002ApJ...577..245J, 2004aap...427..849C}. While almost all of the infrared sources as well as the extended infrared emission of M\,17 is situated to the north and east of the M\,17\,SW molecular gas peak \citep[e.g. in C$^{18}$O;][]{2003ApJ...590..895W}, KW is situated to the southwest of this peak (Figures \ref{fig:all19},\ref{fig:all37}). The more luminous source of the KW binary is named KW-1 and is suspected to be a candidate Herbig Be star \citep{2004aap...427..849C,1997ApJ...489..698H}. Previous model fits to the spectral energy distribution (SED) of KW-1 suggest a $\sim$10\,M$_{\sun}$ central star \citep{2004aap...427..849C, 2009ApJ...696.1278P}, consistent with a B0-B1 ZAMS spectral type \citep{2000AJ....119.1860B}.

Though the KW object is hypothesized to be the central member of a young stellar cluster whose members are all visible in the near-infrared \citep{2004aap...427..849C}, in the \textit{SOFIA} mid-infrared images (Figures \ref{fig:m17},\ref{fig:all19},\ref{fig:all37}), KW is separated by more than an arcminute ($\sim$0.6\,pc) in any direction from any other mid-infrared point source, and is fairly isolated from the bulk of the extended infrared dust emission of M\,17. We do not detect a binary at the location of KW, even in deconvolved 20\,$\mu$m images (not shown) which had a final resolution of 1$\farcs$6 (full width at half maximum) which should have been sufficient to at least marginally resolve the binary whose sources are separated by about 1$\farcs$3. Our astrometric accuracy also is not sufficient to know which of the two near-infrared sources is closest to our mid-infrared emission peaks. \citet{2004aap...427..849C} claim that source KW-1 is the dominant source at 2\,$\mu$m and longer, and thus we are likely only sampling emission from KW-1 with \textit{SOFIA}-FORCAST. If this is the case, our SED models and derived parameters for this source can be considered a good approximation for KW-1 only. In that regard, our model fits to the SED containing the \textit{SOFIA} data (Section \ref{sec:data}) do yield an estimated mass of 8\,M$_{\sun}$ for the KW object, consistent with previous estimates and with the source potentially being a Herbig Be object.

Moreover, we detect a faint larger-scale nebulous emission that is extended east-west around the KW Object at 20\,$\mu$m (Figure \ref{fig:kwo}). At 37\,$\mu$m, the extended emission is also dominantly east-west but, further out ($>$5$\arcsec$), it begins to turn-up and extend more to the north-east. This extension to the northeast is readily seen in the  \textit{Herschel} 70\,$\mu$m data (Figure \ref{fig:kwo}c-d).  Given its morphology as a function of wavelength, it is possible that the extended emission from 20--70\,$\mu$m is tracing an outflow cavity from KW (or an unidentified nearby source). As of now, there has not been any successful attempt to map an outflow from this source in any of the typical outflow tracers, however the near-infrared observations of \citet{2012PASJ...64..110C}, show extended emission and a bipolar polarization pattern with a east-west extension which they claim is indicative of an outflow at a position angle of $\sim$90$\arcdeg$ (i.e. at the angle of the extended mid-infrared emission). Furthermore, \citet{2012ApJ...755..152R} detect a compact radio source  within a few arcseconds of the SOFIA infrared peak of the KW Object and postulate the radio source is associated with KW somehow (Figure \ref{fig:kwo}). They state that the radio spectral index of this source ($\alpha\ge0.9$) is consistent with a hypercompact \ion{H}{2} region, however, it is also consistent with the spectral index of ionized outflows or jets \citep{1986ApJ...304..713R, 2016MNRAS.460.1039P}.

\begin{figure*}[htb!]
\epsscale{1.1}
\plotone{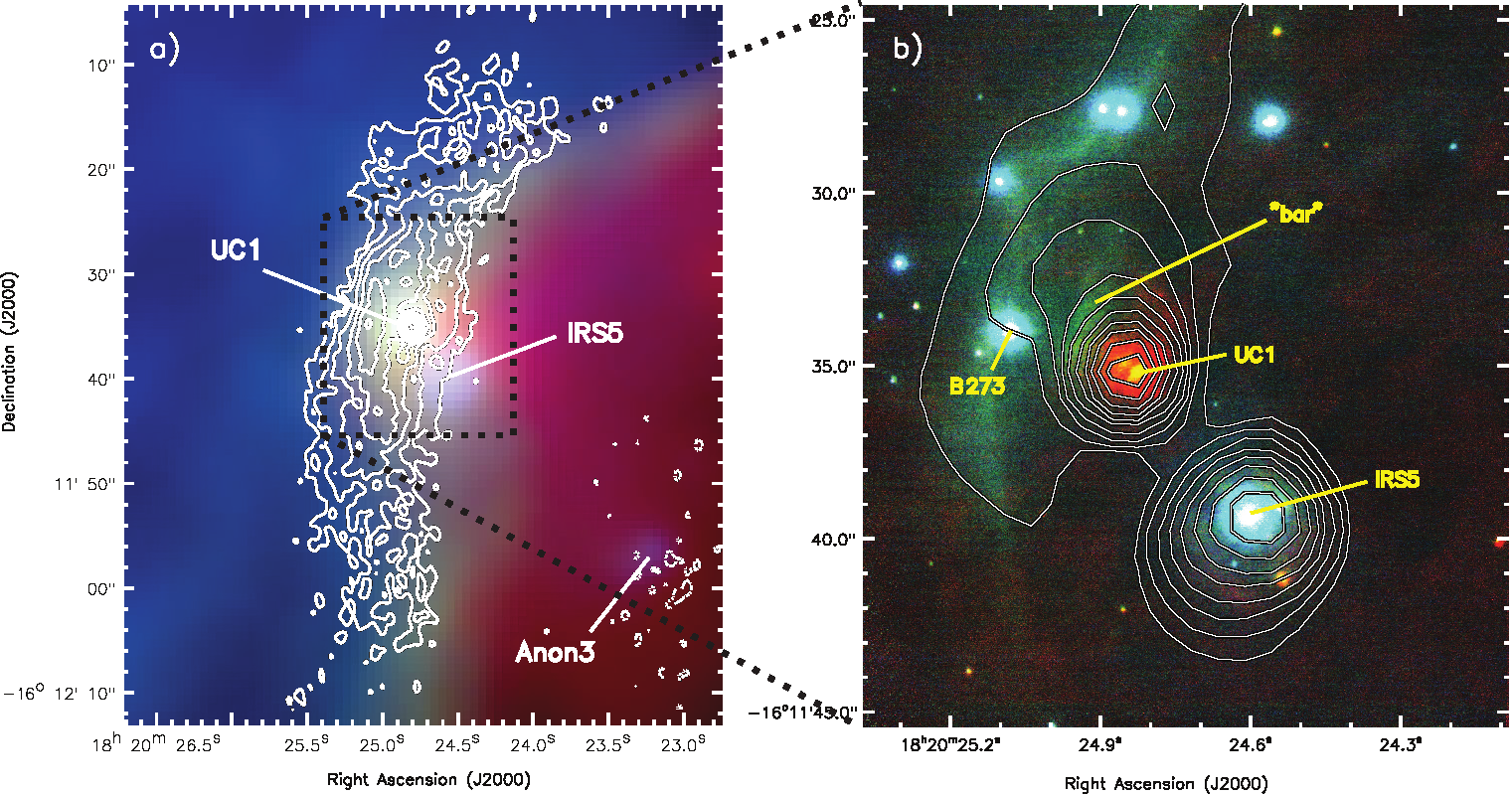}
\caption{UC\,1 and IRS\,5. a) An RGB image (blue; \textit{SOFIA} 20\,$\mu$m; green: \textit{SOFIA} 37\,$\mu$m; red: \textit{Herschel} 70\,$\mu$m) with the 1.3\,cm radio continuum emission from \citet{1998ApJ...500..302J} overlaid as contours. This radio emission shows the bright peak of UC\,1 embedded in the extended emission of the radio ``arc-shaped structure''. b) The deconvolved \textit{SOFIA} 20\,$\mu$m image is shown as contours overlaid onto the near-infrared image (blue: H-band; green: K-band; red: L-band) from \citet{2015AAp..578A..82C}. This region is a zoom-in of the area given by the box in panel a.  \label{fig:uc1b}}
\end{figure*}

\begin{figure*}[htb!]
\epsscale{1.1}
\plotone{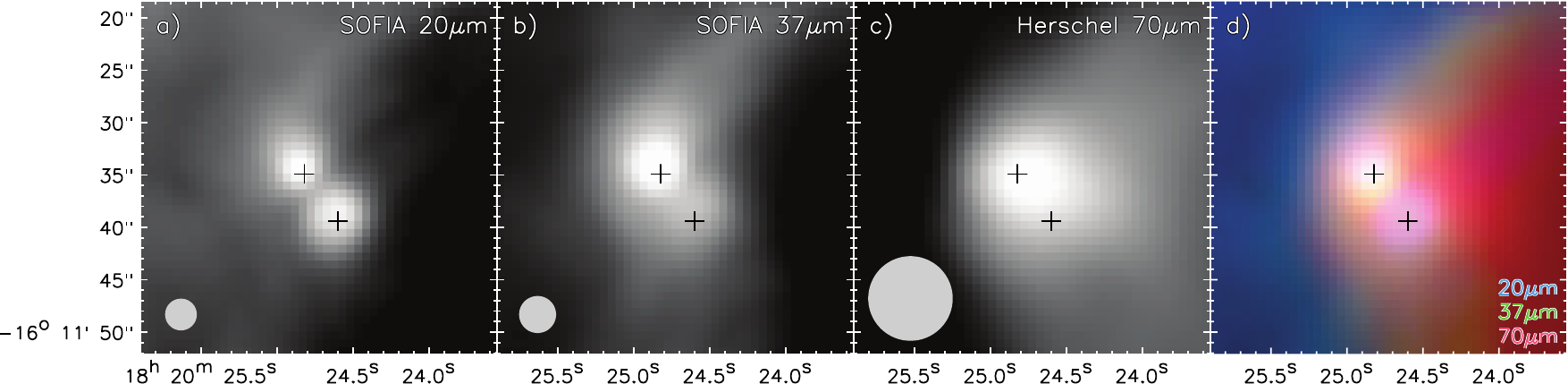}
\caption{UC\,1 and IRS\,5. In each panel the northern plus symbol shows the positions of the near-infrared peak of UC\,1 and the southern plus symbol shows the near-infrared peak of IRS\,5, both from \citet{2015AAp..578A..82C}. a) \textit{SOFIA} 20\,$\mu$m image; b) \textit{SOFIA} 37\,$\mu$m image; and c) \textit{Herschel} 70\,$\mu$m image. The resolution of the images are shown at the lower left corner of each panel. d) An RGB image with the wavelengths representing each color given in the lower right corner of the panel. \label{fig:uc1a}}
\end{figure*}

\subsubsection{The UC~1 and IRS~5 Region}\label{sec:IRS5}
Early radio continuum observations of this region by \citet{1984A&A...136...53F} revealed a bright and compact radio source which was dubbed UC\,1. \citet{1984A&A...136...53F} was also first to resolve the radio continuum emission of UC\,1 into a cometary shape, and given its size ($\sim$0.004\,pc), this source is classified as a hypercompact \ion{H}{2} region \citep{2012ApJ...755..152R}. It is postulated that the central stellar source of UC\,1 is surrounded by an circumstellar accretion disk \citep{2007ApJ...656L..81N}. To the north and south of UC\,1 is an ``arc-shaped structure'' in the radio continuum \citep{1998ApJ...500..302J} which is now believed to be tracing the ionization front between the extended \ion{H}{2} region to the northeast and the molecular cloud of M\,17\,SW to the southwest. IRS\,5 is located $\sim$7$\arcsec$ to the southwest of UC\,1 (Figure \ref{fig:uc1b}) and detected as a bright infrared source with no cm radio continuum emission counterpart \citep{2000A&A...357L..33C}. 

Our 20 and 37\,$\mu$m images of the UC\,1 and IRS\,5 region look very different from each other. Looking at the larger scale environment, there is a bright infrared ridge (labeled ``Ridge'' in Figures \ref{fig:all19} and \ref{fig:all37}) extending for more than an arcminute, with its central portion becoming the radio ``arc-shaped structure'' that curves around UC\,1 and IRS\,5. This infrared ridge is better traced by the 37\,$\mu$m image than the 20\,$\mu$m image; the bulk of the 20\,$\mu$m emission is to the northeast of the ridge, and terminates near the crest of the 37\,$\mu$m emission along the entire ridge. Given that the 20\,$\mu$m filter is very sensitive to the [\ion{S}{3}] line at 18.71\,$\mu$m, it is likely a better tracer of the large-scale \ion{H}{2} emission in the interior of M\,17 rather than the continuum emission from the dusty structures bounding the ionized region. The 20\,$\mu$m emission also traces fairly well the radio continuum and recombination line emission seen by \citet{1998ApJ...500..302J} and the radio ``arc-shaped structure'' (Figure \ref{fig:uc1b}a). Conversely, the 37\,$\mu$m emission, like the emission in the IRAC 3.6--5.8\,$\mu$m images, traces this dusty ridge, however there is some extended 37\,$\mu$m emission to the southwest of the ridge seen only at 37\,$\mu$m.  

\begin{figure*}[htb!]
\epsscale{1.15}
\plotone{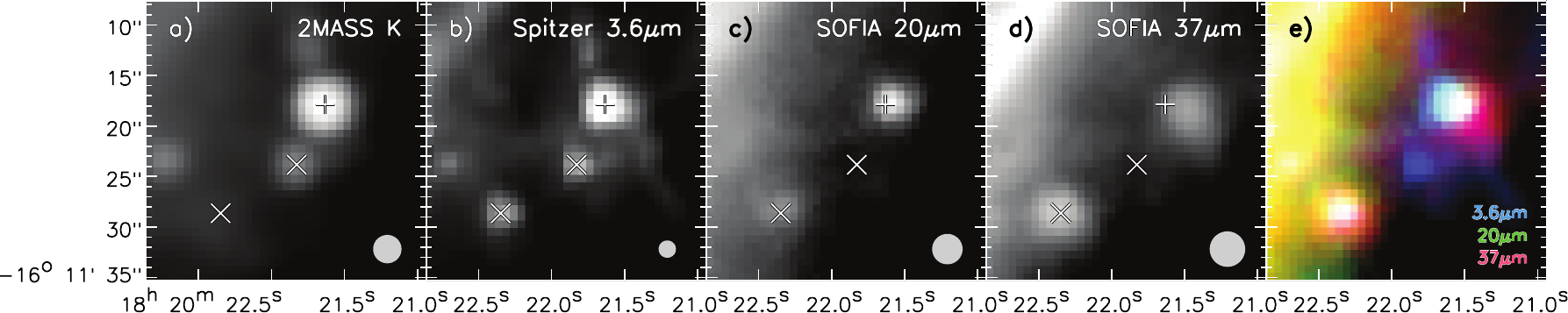}
\caption{CEN\,92. The plus symbol shows the centroid location of CEN\,92 at wavelengths $<$20\,$\mu$m. The centroid location of two astrometric reference sources are shown as X symbols} (the southern most source is Source 5). In panels a-d the wavelength is given in the upper right and the resolution at that wavelength is given by the circle in the lower right. Panel e shows a three-color image with the wavelength of each color given in the lower right. \label{fig:cen92}
\end{figure*}

Now looking at the sources of interest, UC\,1 and IRS\,5 appear to have comparable brightness at 20\,$\mu$m, however at 37\,$\mu$m IRS\,5 is significantly fainter than UC\,1, and in the \textit{Herschel} 70\,$\mu$m images UC\,1 is the only obvious peak in the region (Figure \ref{fig:uc1a}), though there is a high level of environmental emission in the area. For UC\,1, we also detect an extended source that appears elongated to the north at 20\,$\mu$m, but more to the northwest at 37\,$\mu$m. There is no detection of the nearby near-infrared source B\,273 \citep{2015AAp..578A..82C} at these wavelengths (Figure \ref{fig:uc1b}b). \citet{2015AAp..578A..82C} show that at 12\,$\mu$m with high spatial resolution ($\sim$0$\farcs$3) B\,273 is barely detectable, so its emission at 20 and 37\,$\mu$m is probably negligible compared to UC\,1. The northern elongation of UC\,1 at 20\,$\mu$m may be due to the ionization front which extends northward. The extension in emission to the northwest seen at 37\,$\mu$m appears to be following the bright infrared ridge continuing to the northwest from the north of the radio ``arc-shaped structure'' (Figure \ref{fig:uc1b}a). Our derived luminosity ($\sim8.6\times10^5 {L_\sun}$, see \S\ref{sec:cps}) for UC\,1 indicates it is a very massive young source, consistent with the known hypercompact \ion{H}{2} nature of the object.

The overall shape and extent of IRS\,5 looks similar at both 20 and 37$\mu$m. IRS\,5 was shown to be surrounded by four additional, far fainter, near-infrared sources by \citet{2015AAp..578A..82C}. We do not detect/resolve the emission from any of these other nearby sources and at the longer wavelengths of \textit{SOFIA} it appears that the dominant near-infrared source (labeled IRS\,5A by \citealt{2015AAp..578A..82C}) is the only source we are seeing at wavelengths $\ge$20\,$\mu$m. 

While the exact nature of UC\,1 is rather clear, the nature of IRS\,5 is not. Based upon their best-fit models to the SED, \citet{2002AJ....124.1636K} postulate that IRS\,5 is a young B0 star surrounded by a dusty shell in a phase before the onset of an \ion{H}{2} region, and thus at a younger stage of evolution than UC\,1. This does not seem plausible because there is no discernible emission from IRS\,5 in the far-infrared, which is expected from the envelope of a heavily self-embedded pre-ionizing stage of an MYSO. Also, the near-infrared observations by \citet{2015AAp..578A..82C} show emission from IRS\,5 down to wavelengths as short as J-band, while UC\,1 show emission only at wavelengths K-band and longer, signifying that UC\,1 is likely to be more highly embedded than IRS\,5. \citet{2015AAp..578A..82C} postulate that the emission we are seeing in the infrared is perhaps an outflow lobe/cavity, however this also seems unlikely since observations of such structures should show emission in the far-infrared (e.g. \citealt{2017ApJ...843...33D}). The higher spatial resolution ($\sim$0.3$\arcsec$) 11.85\,$\mu$m images from \citet{2015AAp..578A..82C} show that IRS\,5 appears to be an extended region of emission bisected by a dark lane. Given this morphology, if IRS\,5 no longer has an envelope (as evidenced by the lack of far-infrared emission) and has no detectable cm radio continuum emission, we postulate that it is a more evolved Class II YSO with a non-ionizing central star (i.e. has a mass less than $\sim$8\,M$_{\sun}$) with an edge-on disk that is optically thick in the mid-plane in the mid-infrared and where the infrared emission is coming from the flared disk surfaces. Further evidence of the potentially more evolved and lower mass nature of IRS\,5 comes from the near-infrared spectroscopic results of \citet{2015AAp..578A..82C} that show that IRS\,5 does not display hydrogen emission lines that are indicative of ongoing accretion activity, and has a near-infrared spectrum of a mid-/late-B type star. Our SED fitter also cannot find a fit to the data for this source with any of the MYSO models (see \S\ref{sec:data}), again suggesting that IRS\,5 is not a MYSO.

\begin{figure*}[htb!]
\epsscale{0.85}
\plotone{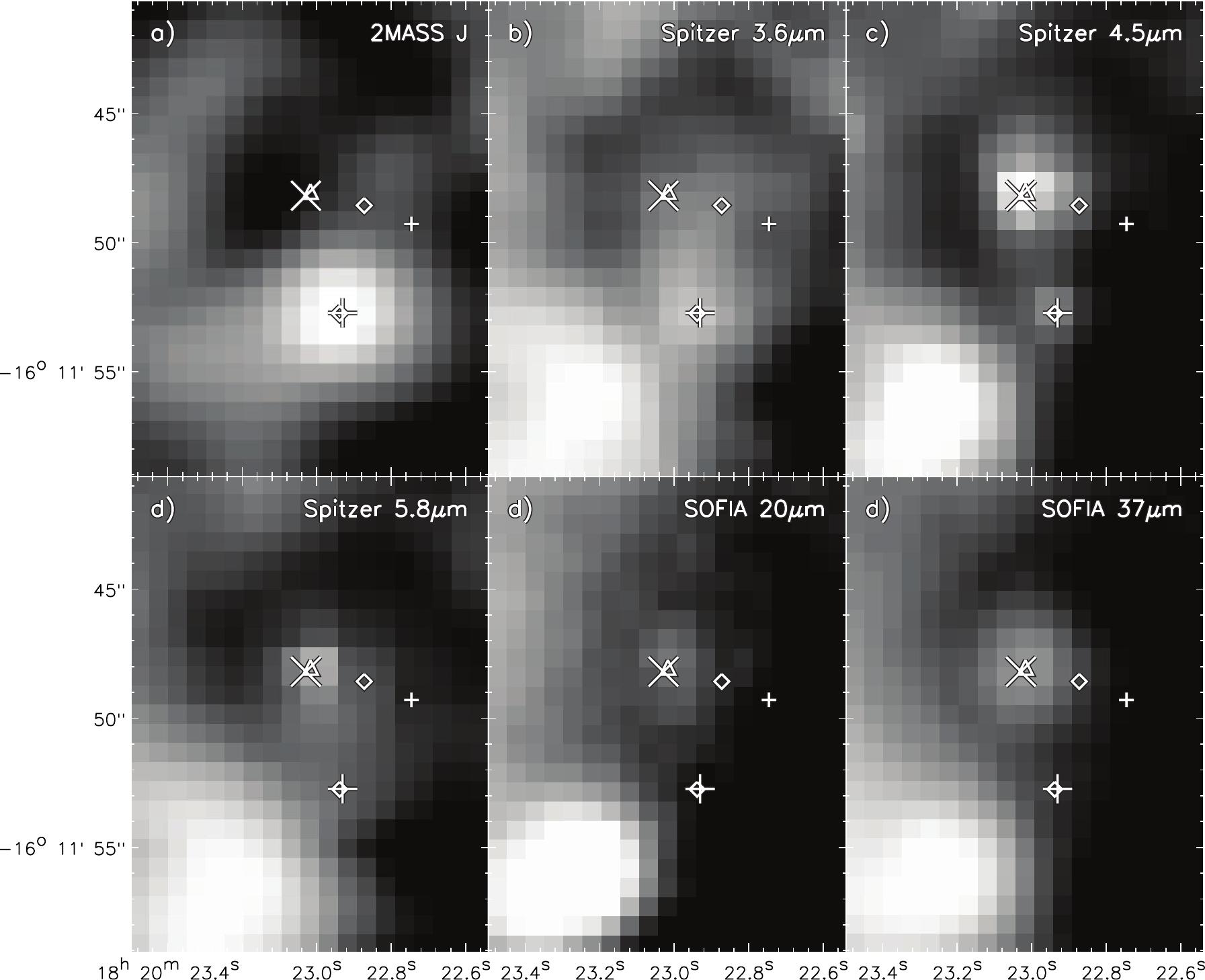}
\caption{Anon1 region from near-infrared to mid-infrared. In each panel the wavelength is given in the upper right and the symbols are: large X symbol is the 37\,$\mu$m peak; the small plus symbol is the location reported by \citet{2001A&A...377..273N} of Anon1; the large plus symbol shows the location of the near-infrared source 2MASS J18202294-1611528; the northern and southern diamond symbols show the positions of the X-ray sources 239 and 244, respectively, from \citet{2007ApJS..169..353B}; and the triangle shows the reference location of the water masers given by \citet{2016MNRAS.460.1839C}. Anon\,3 is the bright source in the lower left corner.  \label{fig:anon1}}
\end{figure*}

\subsubsection{CEN~92 (a.k.a. B~331, IRS~2)}\label{sec:cen92}
CEN\,92 is a source with a known infrared-excess at wavelengths longer than 2\,$\mu$m \citep{2017A&A...604A..78R}, and has been considered to be a MYSO candidate \citep{1997ApJ...489..698H}. Based on optical spectroscopy, \citet{2008ApJ...686..310H} suggest it is a B2 star, consistent with the MYSO hypothesis. \citet{2017A&A...604A..78R} show that CEN\,92 also displays emission line features indicative of accretion from a circumstellar disk, again suggesting that it is a youthful source. However, they also fail to detect helium lines in their near-infrared spectra, which indicates that the source is a late-B or early A-type star. This source is detected at cm radio wavelengths, however, the spectral slope of the radio emission would indicate that it is due to an ionized wind or outflow \citep{2012ApJ...755..152R}, and likely not an ultracompact \ion{H}{2} region. 

Detected by \citet{1998A&A...329..161C} in the optical to near-infrared from U-band to M-band, this source was given the name IRS\,2. This source was also previously observed in the infrared from J-band to Q-band by \citet{2001A&A...377..273N}, and from $\sim$10-20\,$\mu$m by \citet{2002AJ....124.1636K}. Our \textit{SOFIA} observations at 20 and 37\,$\mu$m detect this source at both wavelengths, but it is less prominent at 37 than at 20\,$\mu$m. The emission from this source is peaked at the same location from optical to 5.8\,$\mu$m (i.e. \textit{Spitzer}-IRAC channel 3; the source is not visible due to saturation effects in channel 4), however the 20\,$\mu$m peak is slightly off ($\sim$1$\farcs$5) to the west of this location (Figure \ref{fig:cen92}c), and almost 3$\arcsec$ to the west at 37\,$\mu$m (Figure \ref{fig:cen92}d). Though our absolute astrometric accuracy is roughly 1$\farcs$5, there are several nearby mid-infrared point sources (e.g. Source 5 which is shown for reference in Figure \ref{fig:cen92}, as well as Anon\,1, Anon\,3, and Source 4) which do align to a precision better than this, therefore verifying that these offsets for CEN\,92 at 20 and 37\,$\mu$m are real. At 37\,$\mu$m the emission is cometary shaped with the tail pointing to the southwest. 

With the offsets in emission at 20 and 37\,$\mu$m, it is unclear what the nature of this source is. Given that there is an indication of a radio jet here \citep{2012ApJ...755..152R}, the offset of mid-infrared could be delineating an outflow or outflow cavity. However, these are usually only seen in the mid-infrared if the source is deeply embedded, and CEN\,92 can be readily seen in the optical. Also, in the case of an infrared outflow from a very young YSO, the peak is centered on the stellar source at all wavelengths (if not heavily embedded), or gets closer to the stellar source as one looks at longer wavelengths \citep{2013ApJ...767...58Z, 2017ApJ...843...33D}. This is not what is happening in this case. It is also unlikely that we are seeing two separate YSOs, with the 37\,$\mu$m emission coming from a nearby but more embedded object because no emission is seen coming from either location in images at 70\,$\mu$m or longer wavelengths. It could be that the emission from all wavelengths is from a single source that is a more evolved Class II or III intermediate mass object and the asymmetry of the circumstellar dust is due to photo-evaporation of the eastern side from the NGC\,6618 cluster. Consistent with this, previously-derived values of the luminosity of this source show it to be an intermediate mass object \citep[345\,L$_{\sun}$;][]{2001A&A...377..273N}. Our MYSO fits to the SED fail to fit the data for this source (Section \ref{sec:data}), perhaps due to the lower-mass and/or more evolved nature of this object.

\begin{figure*}[htb!]
\epsscale{0.85}
\plotone{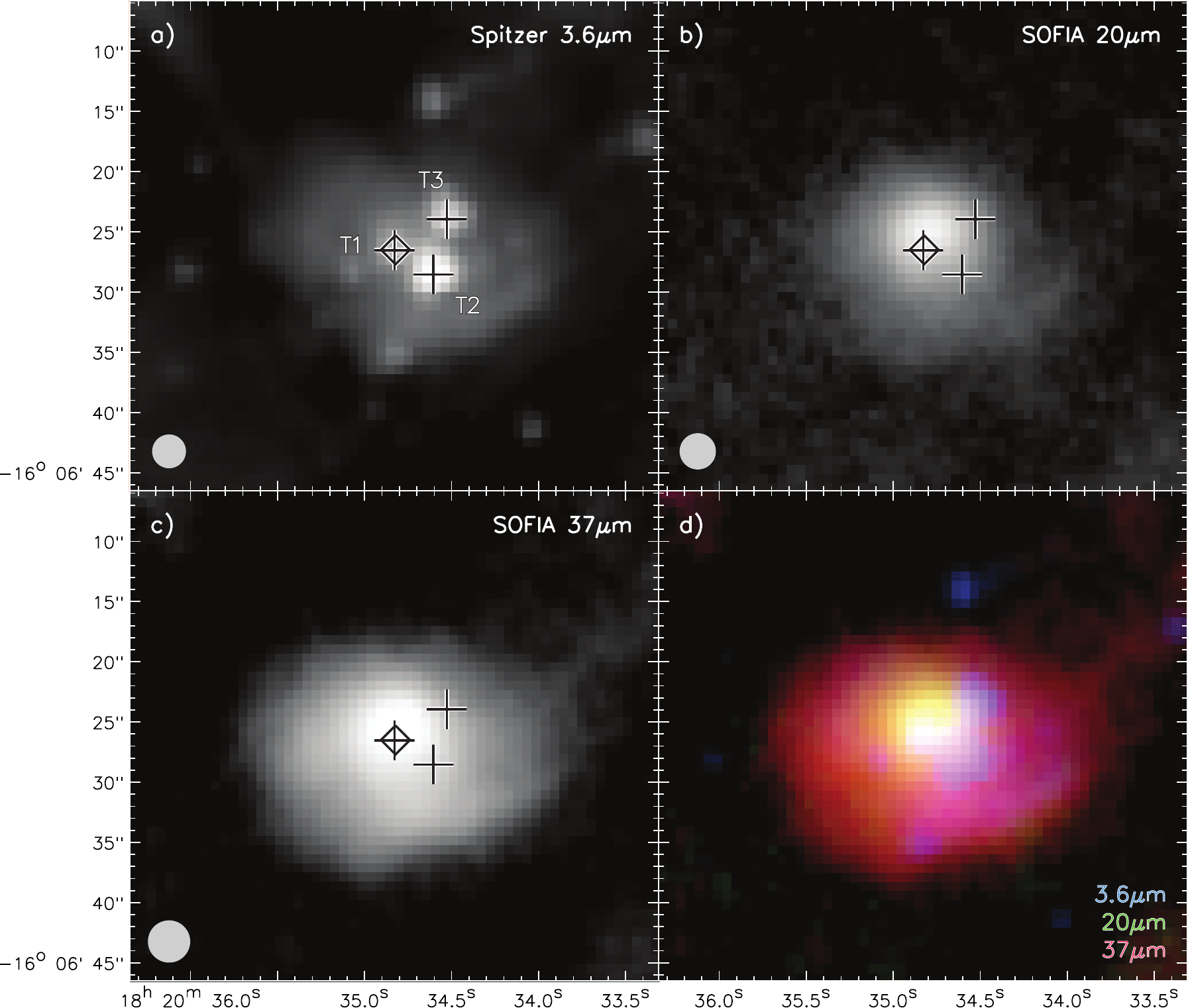}
\caption{G015.128 (G015.1288-00.6717) from near-infrared to mid-infrared. In panels a-c the wavelength is given in the upper right, the diamond is the location of the x-ray source 711 from \citet{2007ApJS..169..353B}, and the large plus signs are the K-band peaks of 2MASS J18203482-160626 (T1), J18203460-1606282 (T2), and J18203452-1606236 (T3). The resolution of the images are shown at the lower left corner of each panel. d) An RGB image with the wavelengths representing each color given in the lower right corner of the panel.\label{fig:trip}}
\end{figure*}

\subsubsection{Anon~1}
Anon\,1 is an extended infrared source first pointed out by \citet{2001A&A...377..273N}, who claim to detect the source at J, K, N, and Q-band. Looking at this region (Figure \ref{fig:anon1}) in the \textit{Spitzer}-IRAC 3.6 to 5.8\,$\mu$m bands, there appears to be two infrared sources close to, but not coincident with, the position of Anon\,1 given by \citet{2001A&A...377..273N}. These two IRAC sources are separated from each other by $\sim$5$\arcsec$, and both are separated from the location of Anon\,1 by about the same amount (Figure \ref{fig:anon1}). The southernmost of the two IRAC sources is the brightest in the 3.6\,$\mu$m image, and is coincident with the \textit{2MASS} source J18202294-1611528 which is prominent in the J, H, and K$_s$ bands. This southern IRAC source is also found in the observations of \citet{2007ApJS..169..353B} to have X-ray emission (source 244), whose properties are that of an unembedded, yet young, star. 

\begin{figure*}[htb!]
\epsscale{1.1}
\plotone{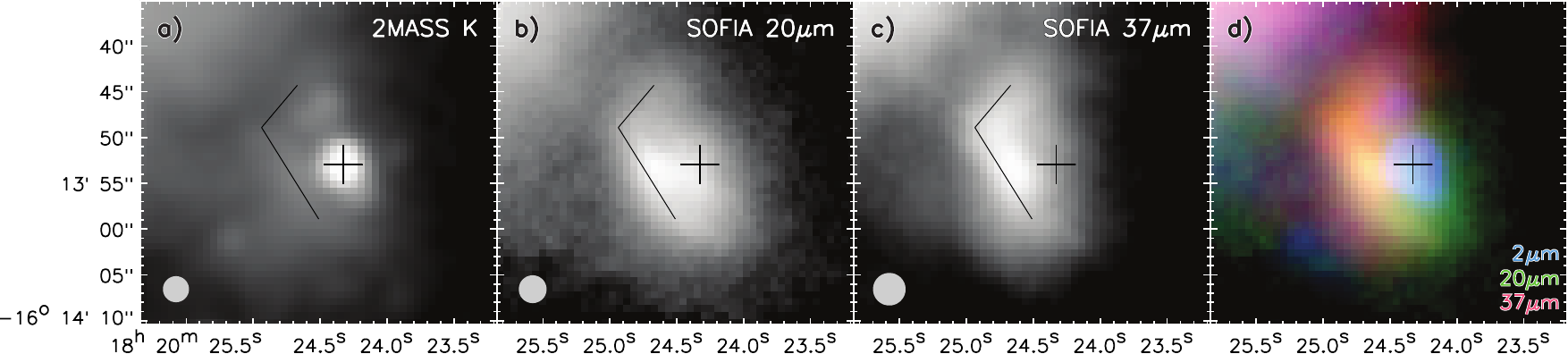}
\caption{Source 7 from near-infrared to mid-infrared. In each panel the wavelength is given in the upper right and the large plus sign is the near-infrared peak of 2MASS J18202432-1613529. The resolution of the images are shown at the lower left corner of each panel. The longer line shows the linear structure to the southeast of 2MASS J18202432-1613529 seen in the near-infrared, and the shorter line shows the location of another linear dust feature seen at longer wavelengths. \label{fig:s7}}
\end{figure*}

Our \textit{SOFIA} 20 and 37\,$\mu$m images of this area show resolved emission of a compact source whose peak is coincident at both wavelengths with the northern-most IRAC source. This source is not seen in the \textit{2MASS} J, H, and Ks data. It is likely, therefore, that Anon\,1 is actually this northern IRAC source which we are also seeing prominently in the \textit{SOFIA} data. That means that the model SED fits by \citet{2001A&A...377..273N} probably employed the J and K-band data of the southern source (2MASS J18202294-1611528), but the N and Q-band data of the northern source. Our SED model fits for Anon\,1 show it to be an intermediate-mass YSO (Section \ref{sec:data}). \citet{1984A&A...136...53F} identify a peak here in the extended 21\,cm radio continuum and claim it is a 0.8\,Jy source, however there is no detection of a source here at 3.5 or 6.0\,cm by \citet{2012ApJ...755..152R} which should have had sufficient sensitivity to detect the source given a reasonable range of source radio spectral indices. Given that our model fits show it to be a likely intermediate mass object, it could be that the radio emission peak of \citet{1984A&A...136...53F} is not tracing Anon\,1. No other peaks identified by \citet{1984A&A...136...53F} are found to  correspond to the infrared sources we see with \textit{SOFIA} (except  UC\,1, of course). However, there are water masers here \citep{1998ApJ...500..302J} coincident with the peak of our mid-infrared emission. \citet{2016MNRAS.460.1839C} resolved the the maser emission and show that the emission comes from discrete spots that appear to reside in an expanding bubble around an embedded YSO. \citet{2007ApJS..169..353B} find an  X-ray source, 239, $\sim$1$\arcsec$ to the west of our mid-infrared peak, and based on the X-ray properties they claim that this emission is most likely tracing a intermediate-mass or high-mass embedded protostellar core, consistent with our SED model fits.

\subsubsection{G015.128 (a.k.a. the Triple)}
G015.1288-00.6717 (or G015.128 for short) was first resolved into three near-infrared sources by \citet{1991ApJ...374..533L}, which are spaced $\sim$2$\farcs$5 ($\sim$5000\,au) from each other. These three sources are referred to as the ``Triple'' by \citet{2002ApJ...577..245J}, who point out that they are surrounded by extended dust emission $\sim$12$\arcsec$ in radius (Figure \ref{fig:trip}). This circumstellar (r$\sim$12$\arcsec$) nebulosity surrounding the triplet was found by  \citet{2012PASJ...64..110C} to show a centro-symmetric polarization pattern in the near-infrared centered on the eastern-most source (their source T1). The emission from this near-infrared nebulosity is elongated at a position angle of 45$\arcdeg$, and they claim that this is an infrared reflection nebula tracing a bipolar outflow driven by T1. Consistent with this hypothesis, the region is found to have water maser emission \citep{2010PASJ...62..391D, 2011MNRAS.418.1689U}, which can be a tracer of outflow activity \citep[e.g.,][]{2005ApJS..156..179D}. 

However, it is the southern-most source (which we will call T2, adopting the labeling by \citealt{2012PASJ...64..110C}; however this source is labeled source 3 by \citealt{2002ApJ...577..245J}) that is the brightest of the three in the near-infrared. T2 was observed in detail by \citet{2017MNRAS.472.3624P}, who claim that the source is likely an embedded A-type supergiant, and due to the presence of certain photospheric absorption lines and high levels of continuum excess, potentially a swollen MYSO. 

In our \textit{SOFIA} images, this region looks like a resolved, single-peaked source (Figure \ref{fig:trip}). Using other nearby stars to confirm our astrometry, we find that the peak seen in both the \textit{SOFIA} 20 and 37\,$\mu$m images is centered closest to the near-infrared source T1.  The distance from the peak position of T1 in the Spitzer 3.6\,$\mu$m image (whose coordinates according to the Spitzer IRAC Handbook have a 0$\farcs$25 error) to the 20\,$\mu$m peak is 1$\farcs$6, and 1$\farcs$0 to the 37\,$\mu$m peak, which means the peaks at all three wavelengths are consistent with being co-spatial to within the combined astrometric errors. There could be smaller emission contribution by T2 at the SOFIA wavelengths since the extended mid-infrared emission pulls in this direction modestly, but the northern-most source of the triplet \citep[T3 from][]{2012PASJ...64..110C} appears to have no significant mid-infrared emission. We conclude that, consistent with \citet{2012PASJ...64..110C}, T1 is likely to be the only MYSO in this region and responsible for the illumination of the infrared nebula seen in the near-infrared. Given the small or non-existent level of emission from the other two sources at wavelengths longer than the near-infrared, those sources are likely to be lower-mass and/or more evolved objects and not MYSOs. Given that the emission is dominated by T1 at longer wavelengths, our modeling of the SED for G015.128 is assumed to be that of T1 only. 

\subsubsection{Source~7}
All sources discussed thus far are of previously identified infrared sources. There are several newly identified mid-infrared sources from this work, however most are simply point sources. Of these newly identified sources, one, Source\,7, does stand out due to the variability in its appearance as a function of wavelength.

Source\,7 is an extended source at 20 and 37\,$\mu$m and is located $\sim$75$\arcsec$ east of KW. However, in the \textit{2MASS} J and H-band images there is a point source here which is catalogued as J18202432-1613529. This source is found to have X-ray emission \citep[source 291;][]{2007ApJS..169..353B} and is considered to be a $\sim$14\,M$_{\sun}$ Class III YSO. At K-band the \textit{2MASS} image starts to show an extended bar-like feature slightly offset to the southeast as well as an extended emission source to the northeast (Figure \ref{fig:s7}). These two structures meet perpendicularly, creating an up-side-down L-shape to the east of J18202432-1613529. This structure and J18202432-1613529 are both seen at 20\,$\mu$m with \textit{SOFIA} (Figure \ref{fig:s7}), however now the L-shaped emission feature is brighter and J18202432-1613529 is fainter. At 37\,$\mu$m, we only see the L-shaped structure. It is unclear what this bar is, and how it is related to the near-infrared source. In \citetalias{2019ApJ...873...51L}, we saw compact infrared sources interior to dusty arcs (and even multiple nested arcs), and this might be an analogous type of structure. In our SED analysis (Section \ref{sec:data}), we assume that the emission is related to and heated by the stellar source and we derive a best-fit model with a mass of 8\,M$_{\sun}$, which is slightly less than the mass estimate in \citet{2007ApJS..169..353B}.

\begin{deluxetable*}{rrrrrrrrrrrr}
\tabletypesize{\scriptsize}
\tablecolumns{8}
\tablewidth{0pt}
\tablecaption{Observational Parameters of Compact Sources in M\,17}
\tablehead{\colhead{  }&
           \colhead{  }&
           \colhead{  }&
           \multicolumn{3}{c}{${\rm 20\mu{m}}$}&
           \multicolumn{3}{c}{${\rm 37\mu{m}}$}&
           \colhead{  }\\
           \cmidrule(lr){4-6} \cmidrule(lr){7-9} \\
           \colhead{ Source }&
           \colhead{ R.A. } &
           \colhead{ Dec. } &
           \colhead{ $R_{\rm int}$ } &
           \colhead{ $F_{\rm int}$ } &
           \colhead{ $F_{\rm int-bg}$ } &
           \colhead{ $R_{\rm int}$ } &
           \colhead{ $F_{\rm int}$ } &
           \colhead{ $F_{\rm int-bg}$ } &
           \colhead{Aliases}\\
	   \colhead{  } &
	   \colhead{  } &
	   \colhead{  } &
	   \colhead{ ($\arcsec$) } &
	   \colhead{ (Jy) } &
	   \colhead{ (Jy) } &
	   \colhead{ ($\arcsec$) } &
	   \colhead{ (Jy) } &
	   \colhead{ (Jy) } &
	   \colhead{} 
}
\startdata
         KW & 18 20 19.4 &-16 13 29.5 &  7.68 &    92.86 &   92.705 & 15.36 &   322.33 &   306.73 &  M17SW-IRS1\\ 
       IRS5 & 18 20 24.6 &-16 11 39.4 &  3.84 &   236.85 &   172.48 &  3.84 &   590.96 &   383.52 &  \\ 
        UC1 & 18 20 24.8 &-16 11 34.3 &  3.84 &   272.31 &   191.25 &  3.84 &  1285.05 &  1081.54 &  \\ 
      CEN92 & 18 20 21.6 &-16 11 17.7 &  3.84 &   22.67 &     7.90 &  4.61 &   128.98 &    21.89 &  B331, IRS2\\ 
      Anon1 & 18 20 23.0 &-16 11 47.6 &  3.84 &   19.46 &     0.29 &  3.84 &    84.42 &     5.75 &  \\ 
      Anon3 & 18 20 23.3 &-16 11 56.1 &  5.38 &   53.68 &    17.96 &  5.38 &   233.41 &    61.57 &  \\ 
   G015.128 & 18 20 34.8 &-16 06 24.1 & 13.06 &   112.02 &    64.09 & 15.36 &   554.17 &   394.59 &  Triple\\ 
          1 & 18 20 15.7 &-16 14 09.8 &  6.91 &   40.62 &    37.01 & 11.52 &   125.35 &   121.64 &  \\ 
          2 & 18 20 18.7 &-16 11 57.3 &  3.84 &   2.61 &     0.93 &  3.84 &    40.61 &     7.80 &  \\ 
          3 & 18 20 19.3 &-16 12 05.9 &  3.84 &   1.41 &     0.36 &  3.84 &    35.88 &     6.74 &  \\ 
          4 & 18 20 19.6 &-16 10 38.1 &  3.07 &   4.70 &     0.93 &  2.30 &    22.37 &     1.31 &  \\ 
          5 & 18 20 22.3 &-16 11 28.3 &  3.84 &   22.15 &     2.90 &  3.84 &   104.40 &    21.66 &  \\ 
          6 & 18 20 22.8 &-16 13 48.0 &  6.14 &   8.11 &     3.92 &  7.68 &    75.38 &    24.64 &  \\ 
          7 & 18 20 24.6 &-16 13 54.0 & 11.52 &   118.72 &    66.57 & 13.06 &   527.99 &   266.93 &  \\ 
          8 & 18 20 29.1 &-16 06 00.7 &  3.84 &   5.82 &     3.34 &  3.84 &     9.74 &     4.93 &  \\ 
          9 & 18 20 37.6 &-16 10 02.6 &  3.07 &   3.42 &     0.74 &  4.61 &    23.91 &    15.30 &  \\ 
\enddata
\tablecomments{\footnotesize R.A. and Dec. are for the center of the photometric apertures used. $F_{\rm int}$ indicates total flux inside the aperture. $F_{\rm int-bg}$ is for background subtracted flux. The estimated photometric uncertainties are 15\% for 20\,$\mu$m and 10\% for 37\,$\mu$m.}
\label{tb:cps1}
\end{deluxetable*}

\begin{figure}
\epsscale{1.0}
\plotone{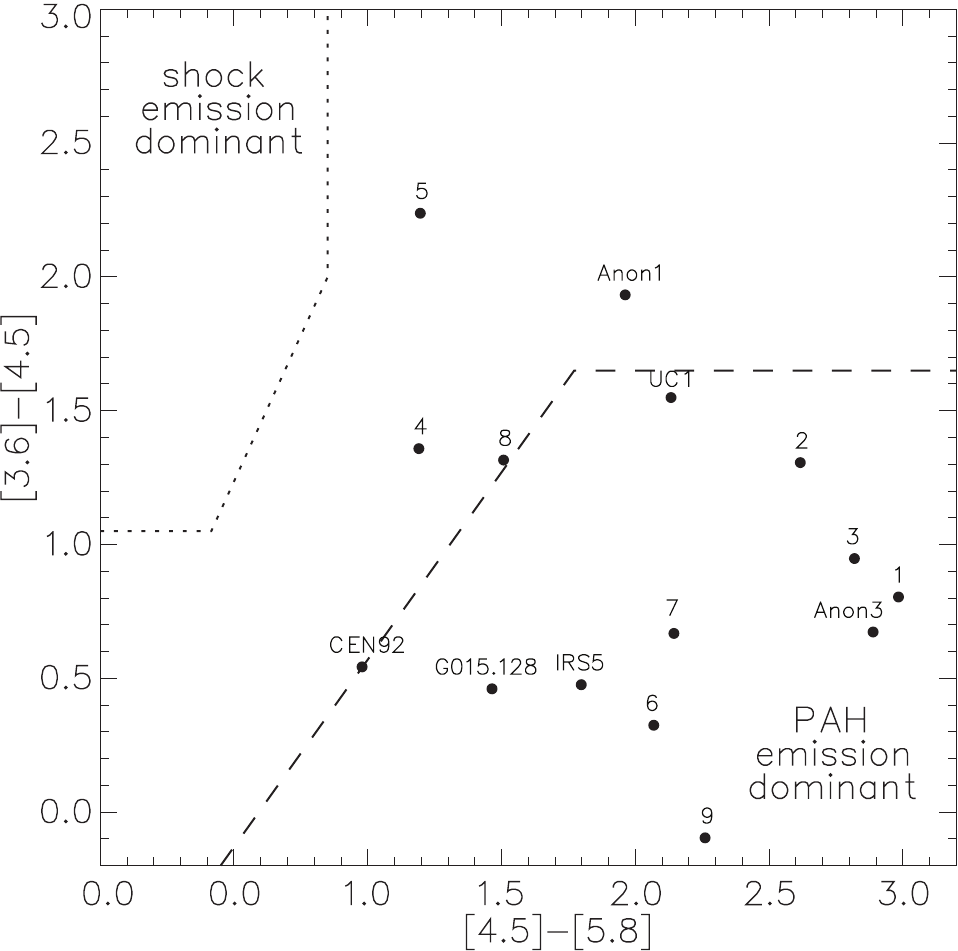}
\caption{\footnotesize A color-color diagram utilizing our background-subtracted \textit{Spitzer}-IRAC 3.6, 4.5, 5.8 and 8\,$\mu$m source photometry to distinguish ``shocked emission dominant'' and ``PAH emission dominant'' YSO candidates from our list of sub-components and point sources. Above (up-left) of dotted line indicates shock emission dominant regime. Below (bottom-right) dashed line indicates PAH dominant regime. We adopt this metric from \citet{2009ApJS..184...18G}.}
\label{fig:ccd}
\end{figure}

\section{Data Analysis and Results}\label{sec:data}

\subsection{Physical Properties of Sub-Components and Point Sources: SED Model Fitting and Determining MYSO Candidates}\label{sec:cps}

Using the source-finding algorithm from \citetalias{2019ApJ...873...51L} \citep[i.e. an optimal extraction method;][]{1998MNRAS.296..339N}, we identified all potential point-like or compact sources, and have compiled a list in Table \ref{tb:cps1} consisting of those sources that are either isolated from the extended infrared emission of the M\,17 nebula or are visible as peaks at both 20 and 37\,$\mu$m. Among these 16 objects in total are the seven individual sources discussed in Section \ref{sec:sources}. We also include Source 1 for which we have photometric data, though the region containing this source was not included in the maps in Figures \ref{fig:all19} and \ref{fig:all37} due to uncorrectable spatial distortion issues in the images (see discussion in Section \ref{sec:obs}). Source 1 is believed to be spatially coincident with the ammonia core MSX6C G014.9790-00.6649 \citep{2011MNRAS.418.1689U} and that position is labeled in Figures \ref{fig:all19} and \ref{fig:all37}. 

Table \ref{tb:cps1} contains the information regarding the position, radius employed for aperture photometry, and 20 and 37\,$\mu$m flux densities (before and after background subtraction) of all these sources. In addition to using the photometry from the \textit{SOFIA} data, we performed multi-band aperture photometry on the \textit{Spitzer}-IRAC 3.6, 4.5, 5.8, 8.0\,$\mu$m (Table \ref{tb:cps2}) and \textit{Herschel}-PACS 70 and 160\,$\mu$m image data (Table \ref{tb:cps3}) on all sources (see Appendix \ref{appendixc}). We employed the same optimal extraction technique as in \citetalias{2019ApJ...873...51L} to find the optimal aperture to use for photometry. Background subtraction was also performed in the same way as \citetalias{2019ApJ...873...51L} (i.e. using background statistics from an annulus outside the optimal extraction radius which had the least environmental contamination).  

We used these data to create the near-infrared to far-infrared SEDs of the identified sources with the intent to fit them with SED models of MYSOs \citep{2011ApJ...733...55Z}. In the case of KW, the \textit{Spitzer}-IRAC data are saturated at all wavelengths, so we used the compiled flux measurements from \citet{2004aap...427..849C} to assist in creating a complete SED for this source (see Table\,\ref{tb:kw}). We also adopted the flux values of IRS\,5, and UC\,1 from \citet{2002AJ....124.1636K} and \citet{2015AAp..578A..82C}. Those bands fill the wavelength gaps between \textit{Spitzer}-IRAC and \textit{SOFIA}-FORCAST bands. Additionally, we used the infrared flux values from the \citet{2002AJ....124.1636K} for CEN\,92 (see Table\,\ref{tb:others}).

\begin{deluxetable*}{rcccccrclrcll}
\tabletypesize{\scriptsize}
\tablecolumns{12}
\tablewidth{0pt}
\tablecaption{SED Fitting Parameters for Compact Sources in M\,17}
\tablehead{\colhead{   Source   } &
           \colhead{  $L_{\rm obs}$   } &
           \colhead{  $L_{\rm tot}$   } &
           \colhead{ $A_v$ } &
           \colhead{  $M_{\rm star}$  } &
           \colhead{ Best } &
           \multicolumn{3}{c}{$A_v$ Range}&
           \multicolumn{3}{c}{$M_{\rm star}$ Range}&
           \colhead{ Notes }\\
	   \colhead{        } &
	   \colhead{ ($\times 10^3 L_{\sun}$) } &
	   \colhead{ ($\times 10^3 L_{\sun}$) } &
	   \colhead{ (mag.) } &
	   \colhead{ ($M_{\sun}$) } &
       \colhead{  Models\tablenotemark{a} } &
       \multicolumn{3}{c}{(mag.)}&
       \multicolumn{3}{c}{($M_{\sun}$)}&
       \colhead{     }\\
}
\startdata
             KW &      4.38 &      9.48 &     31.9 &      8.0 & 7  &  1.7 & - &  50.3 &   8.0 & - &  24.0 & MYSO; 3.5cm (HCHII/jet)\\
           IRS5 &      1.19 &    157.86\tablenotemark{b} &     25.2\tablenotemark{b} &     32.0\tablenotemark{b} & 5  & 21.2 & - &  33.5\tablenotemark{b} &  24.0 & - &  32.0\tablenotemark{b} & \\
            UC1 &     22.96 &    858.44 &     79.5 &     64.0 & 8  & 53.0 & - &  79.5 &  48.0 & - &  64.0 & MYSO; 1.3,3.5,6.0cm (UCHII)\\
          CEN92 &      0.14 &     12.32\tablenotemark{b} &      5.3\tablenotemark{b} &     12.0\tablenotemark{b} & 5  &  2.5 & - &  26.0\tablenotemark{b} &   0.5 & - &  16.0\tablenotemark{b} & 3.5,6.0cm (jet)\\
          Anon1 &      0.14 &      0.68 &     76.3 &      4.0 & 7  & 10.9 & - &  76.3 &   2.0 & - &   4.0 & \\
          Anon3 &      1.07 &      2.59 &     26.5 &      2.0 & 5  & 26.5 & - &  75.5 &   2.0 & - &  24.0 & \\       
       G015.128 &      6.19 &     49.40 &     26.5 &     12.0 & 10  & 1.7 & - &  26.5 &   8.0 & - &  12.0 & MYSO\\
              1 &      3.01 &     13.30 &     26.5 &      8.0 & 10  & 8.4 & - &  31.0 &   8.0 & - &  16.0 & MYSO\\
              2 &      0.22 &      0.31 &     36.9 &      2.0 & 9  & 14.3 & - &  36.9 &   2.0 & - &   2.0 & \\
              3 &      0.13 &      0.68 &     77.1 &      4.0 & 13  & 16.8 & - & 244.0 &   2.0 & - &  96.0 & \\
              4 &      0.08 &      0.67 &     42.4 &      4.0 & 12  & 10.6 & - &  58.7 &   0.5 & - &  12.0 & \\
              5 &      0.27 &      0.76 &    100.6 &      2.0 & 7  & 2.5 & - & 101.0 &   2.0 & - &  16.0 & \\
              6 &      0.54 &      0.79 &      5.0 &      4.0 & 7  & 1.7 & - &  26.8 &   4.0 & - &   4.0 & \\
              7 &      4.76 &      9.67 &      9.2 &      8.0 & 8  & 8.4 & - &  53.0 &   8.0 & - &   8.0 & MYSO\\
              8 &      0.21 &     10.84 &     73.8 &     12.0 & 12  & 59.5 & - &  82.2 &  12.0 & - &  12.0 & MYSO\\
              9 &      0.40 &    157.86 &    184.4 &     32.0 & 9 & 49.5 & - & 210.0 &   4.0 & - &  32.0 & pMYSO\\
\enddata
\tablenotetext{a}{\scriptsize The number of models in the group of best fit models. These models were used to determine the ranges of $M_{\rm star}$ and $A_v$.}
\tablenotetext{b}{\scriptsize These sources are not considered to be MYSOs due to the fact that the SED fits to the data for these sources are poor. Further information in $\S$\ref{sec:IRS5} and $\S$\ref{sec:cen92} lead us to believe they are both perhaps intermediate mass Class II sources.}
\tablecomments{\scriptsize A ``MYSO'' in the right column denotes a MYSO candidate. A ``pMYSO'' indicates that their is greater uncertainty in the derived physical parameters and that these sources are possible MYSO candidates. A detailed description of these definitions is given in Section 4.1 of \citetalias{2019ApJ...873...51L}.  If the infrared source is a point source in cm radio continuum, or at the location of a prominent radio continuum peak, the wavelength of this is given in the right column \citep[data are from][]{2012ApJ...755..152R}, along with any identification of the nature of the radio emission [HCHII: hypercompact HII region; UCHII: ultracompact HII region; jet (based on spectral radio index)].}
\label{tb:sedp}
\end{deluxetable*}

An issue with the fluxes derived from the \textit{Spitzer}-IRAC data of YSOs at 3.6, 5.8 and 8\,$\mu$m is that they can be contaminated by PAH emission \citep{2001ApJ...548L..73H, 2007ApJ...657..810D}, and the 4.5\,$\mu$m fluxes can be contaminated by shock-excited H$_2$ emission \citep{2006AJ....131.1479R}. As discussed in \citetalias{2019ApJ...873...51L}, a color-color diagram ([3.6]-[4.5] vs. [4.5]-[5.8]) can be used to determine if sources are highly contaminated by shock emission and/or PAH emission, and we have employed that technique here again for the sources in M\,17 (Figure \ref{fig:ccd}). Contaminated IRAC fluxes are set as upper limits to the photometry used in constructing the SEDs (and later in the SED fitting). Note that the IRAC color-color diagram shows that the IRS\,5 and UC\,1 sources are contaminated by PAH emission. To be consistent with this, we also set the 8.7\,$\micron$ data that we adopted from the literature as upper limits for these sources due to the likelihood of those flux values being contaminated by the 8.6\,$\micron$ PAH feature. Though we could not put KW on the color-color diagram due to saturation issues, we will make this same assumption for the 8.7\,$\micron$ data that we adopted from the literature for KW.   

The \textit{Herschel} 70 and 160\,$\mu$m data are set to be upper limits for most sources due to the coarser spatial resolution ($\sim$10$\arcsec$) of the data and the high likelihood that the photometry is contaminated by emission from adjacent sources or the extended dusty environment of M\,17. However, the \textit{Herschel} 70\,$\micron$ data of UC\,1, KW, G015.128, source 1, and source 9 are treated as nominal data points since they are isolated from or can be easily distinguished from any environmental contamination (i.e., they had relatively flat background profiles which could be properly subtracted by optimal extraction). For this same reason, the \textit{Herschel} 160\,$\micron$ data for source 9 is employed as nominal data point.

In general, we expect that MYSO fluxes will increase as a function of wavelength in the near-infrared to far-infrared. \citet{2015AAp..578A..82C} provided fluxes for UC\,1 and IRS\,5 at 17.72\,$\micron$, and as expected they are less than our SOFIA 20\,$\micron$ fluxes (though they also agree to with within the combined photometric errors). However, the 20.6\,$\micron$ flux values of \citet{2002AJ....124.1636K} for these two sources are unexpectedly less than the 17.72\,$\micron$ fluxes, and quite significantly less than our 20\,$\micron$ flux values. \citet{2002AJ....124.1636K} also provide a 20.6\,$\micron$ flux value for CEN\,92 which is also slightly less than our SOFIA 20\,$\micron$ flux. Given this inconsistency with regards to the expected flux vs. wavelength behavior of MYSOs, and given the systematically lower flux values for these sources compared to our data, we do not include the 20.6\,$\micron$ flux values from \citet{2002AJ....124.1636K} in our analyses of IRS\,5, UC\,1, and CEN\,92.

As we did in \citetalias{2019ApJ...873...51L}, we set the upper error bar on our photometry as the subtracted background flux value (since background subtraction can be highly variable but never larger than the amount being subtracted), and the lower error bar values for all sources come from the average total photometric error at each wavelength (as discussed in Section \ref{sec:obs} and \citetalias{2019ApJ...873...51L}) which are set to be the estimated photometric errors of 20\%, 15\%, and 10\% for 4.5, 20, and 37\,$\mu$m bands, respectively. We assume that the photometric errors of the {\it Spitzer}-IRAC 3.6, 5.8, and 8.0\,$\micron$ fluxes are 20\,\% for the sources that are not contaminated by PAH features. The lower error bars of the 70 and 160$\micron$ data points are assumed to be 40$\%$ and 30$\%$, respectively, adopting the most conservative (largest) uncertainties of the {\it Herschel} compact source catalog \citep{2016A&A...591A.149M,2017MNRAS.471..100E}. We also consider additional uncertainties for the \textit{SOFIA} 20 and 37\,$\micron$ photometry of KW since it was located at an area in the map where we combined the data from a flight which suffered from poorer flux calibration accuracy (see discussion in Section \ref{sec:obs}). Therefore, for KW only, we assume larger total uncertainties of $20\%$ and $15\%$ for the 20 and 37$\micron$ photometry, respectively.

\begin{figure*}[htb!]
\plotone{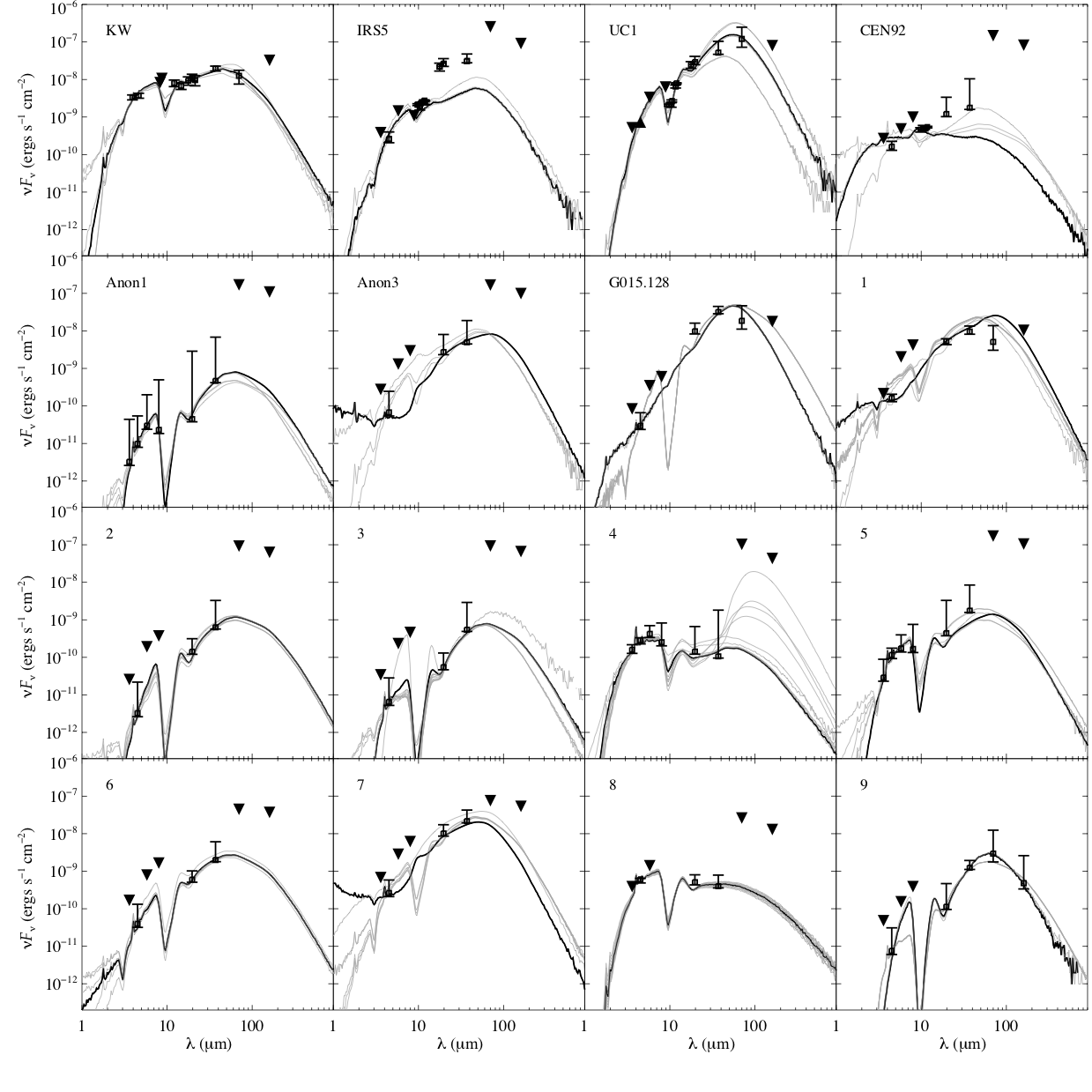}
\caption{SED fitting with ZT model for compact sources in M\,17. Black lines are the best fit model to the SEDs, and the system of gray lines are the remaining fits in the group of best fits (from Table\,\ref{tb:sedp}). Upside-down triangles are data that are used as upper limits in the SED fits. For the discussion of error bars and why some data are used as upper limits while others are not for different sources, see \S\,\ref{sec:cps}.}
\label{fig:fd1}
\end{figure*}

Once SEDs could be constructed from the photometric data (and their associated errors or limits), we utilized the ZT \citep{2011ApJ...733...55Z} MYSO SED model fitter as we did in \citetalias{2019ApJ...873...51L} in order to investigate the physical properties of individual sources. The fitter pursues a $\chi^2$-minimization to determine the best fit MYSO model and provides additional models ordered by their $\chi_{\rm nonlimit}^2$ values. Being consistent with the analysis of \citetalias{2019ApJ...873...51L}, we select a group of models that show $\chi_{\rm nonlimit}^2$ values similar to the best fit model and distinguishable from the next group of models showing significantly larger $\chi_{\rm nonlimit}^2$ values (see further discussion in \citetalias{2019ApJ...873...51L}).

Figure~\ref{fig:fd1} shows the ZT MYSO SED model fits as the solid lines (black for the best model fit, and gray for the rest in the group of best fit models) on top of the derived photometry points for each individual source. Table~\ref{tb:sedp} lists the physical properties of the MYSO SED model fits for each source. The observed bolometric luminosities, $L_{\rm obs}$, of the best fit models are presented in column~2 and the true total bolometric luminosities, $L_{\rm tot}$ (i.e. corrected for the foreground extinction and outflow viewing angles), in column~3. The extinction and the stellar mass of the best models are listed in column~4 and 5, respectively. In column~6, we provide the number of the models in the group of best model fits. The column~7 and 8 present the ranges of the foreground extinction and stellar masses derived from the models in the group of best model fits in column~6. Column~9 shows the identification of the individual sources based on the previous studies as well as our criteria of MYSOs and possible MYSOs (``pMYSOs'') set in \citetalias{2019ApJ...873...51L}. To summarize, the conditions for a source to be considered a MYSO is that it must 1) have an SED reasonably fit by the MYSO models, 2) have a M$_{\rm star}\ge8\,$M$_{\sun}$ for the best model fit model, and 3) have M$_{\rm star}\ge8\,$M$_{\sun}$ for the range of $M_{\rm star}$ of the group of best fit models. A pMYSO fulfills only the first two of these criteria. 

Looking to the inventory of compact sources and their observational characteristics and derived physical parameters in M\,17, we can make some overall assessments as well as gross comparisons to the results of W\,51\,A from \citetalias{2019ApJ...873...51L}. Though the number of identified sources is a third the number found in W\,51\,A, we found that there are no shock-dominated sources identified in our M\,17 sample (W\,51\,A had only one such source), and 69\% of the sources in M\,17 are PAH-dominated which is comparable to the 77\% seen in W\,51\,A. Perhaps more striking is that for W\,51\,A, 41 of the 47 compact sources (87\%) were found via SED fitting to likely be MYSOs, whereas for M\,17 only 7 of 16 sources (44\%) appear to be MYSOs. One likely reason for this difference is that M\,17 is almost three times closer than W\,51\,A, and many of the lower-mass objects being detected in M\,17 would not be detectable if they were at the distance of W\,51\,A. The most massive source in the M\,17 region by far is UC\,1, weighing in at a best fit mass of 64\,M$_{\sun}$. While this is a sizable source, by comparison, we found in \citetalias{2019ApJ...873...51L} that W\,51\,A contains 8 MYSOs with best-fit masses equal to or greater than 64\,M$_{\sun}$.

There are only two sources that are not fit well by the MYSO fitting algorithm we employ here. Looking at both IRS\,5 and CEN\,92 in Figure\,\ref{fig:fd1}, we see that the data points are not well fit by any of the models to within their error bars. For both of these sources, as we discussed in Section\,\ref{sec:sources}, there are peculiarities in their observational properties that lead us to assume they are not embedded MYSOs, but instead perhaps lower-mass, non-ionizing Class II objects. If these sources are indeed low- or intermediate-mass Class II sources, then this may be the reason why they are not well-fit by the MYSO fitter. 

\begin{deluxetable*}{rrrrrrrrrr}
\centering
\tabletypesize{\scriptsize}
\tablecolumns{8}
\tablewidth{0pt}
\tablecaption{Observational Parameters of Major Extended Sources in M\,17}
\tablehead{\colhead{  Source  }&
           \colhead{   R.A.  } &
           \colhead{   Dec.  } &
           \colhead{ $a_{\rm tot}$ } &
           \colhead{ $b_{\rm tot}$ } &
           \colhead{ $PA$ } &
           \colhead{      $F_{\rm 20\micron,tot}$      } &
           \colhead{      $F_{\rm 37\micron,tot}$      } \\
	   \colhead{  } &
	   \colhead{  } &
	   \colhead{  } &
	   \colhead{ ($\arcsec$) } &
	   \colhead{ ($\arcsec$) } &
	   \colhead{ ($\degr$) } &
	   \colhead{ (Jy) } &
	   \colhead{ (Jy) } &
}
\startdata
M\,17\,N & 18 20 34.2 & -16 08 37.7 & 288.40  & 103.96  & 25      &  2.27E+04 & 5.68E+04  \\
M\,17\,S & 18 20 23.8 & -16 12 22.8 & 218.99  & 100.51  & 50      &  2.76E+04 & 8.21E+04  \\
Triple    & 18 20 35.2 & -16 06 12.0 & 34.06   & \nodata & \nodata &  3.08E+02 & 1.17E+03  \\
Cavity    & 18 20 31.2 & -16 10 54.8 & 39.30   & \nodata & \nodata &  4.67E+02 & 1.79E+03  \\
\enddata
\tablecomments{\footnotesize R.A. and Dec. are for the center of the ellipses which have semi-major and semi-minor axes defined by  \textit{$a_{\rm tot}$} and \textit{$b_{\rm tot}$}, respectively. The fluxes for the sources Cavity and Triple are defined by circular aperture whose radii are given solely by \textit{$a_{\rm tot}$}. Photometric uncertainties are estimated to be 40\% for both 20\,$\mu$m and 37\,$\mu$m.}
\label{tb:es}
\end{deluxetable*}

\begin{deluxetable*}{rrrcrrrrrrrr}
\centering
\tabletypesize{\scriptsize}
\tablecolumns{12}
\tablewidth{0pt}
\tablecaption{Derived Parameters of Major Extended Sources in M\,17}
\tablehead{\colhead{  Source  }&
	   \colhead{   $M_{\rm vir}$  } &
           \colhead{   $M$  } &
           \colhead{   $L$  } &
           \colhead{   $T_{\rm cold}$  } &
           \colhead{   $T_{\rm warm}$  } &
           \colhead{   $L/M$  } &
           \colhead{ $\alpha_{\rm vir}$ } \\
	   \colhead{            } &
	   \colhead{   ($M_{\sun}$)   } &
	   \colhead{   ($M_{\sun}$)   } &
	   \colhead{   ($\times10^{4}L_{\sun}$)   } &
	   \colhead{   ($K$)   } &
	   \colhead{   ($K$)   } &
	   \colhead{   ($L_{\sun}/M_{\sun}$)   } &
	   \colhead{             }  }
\startdata
M\,17\,N   & 772.8  & 481.4   & 195  & 47.7 & 186.1 & 2022.8 &  1.61\\
M\,17\,S   & 1261.8 & 4337.2  & 296  & 50.1 & 188.6 &  340.9 &  0.29\\
Triple      & 338.9  & 48.3    & 3.01 & 46.8 & 203.2 &  312.1 &  7.02\\
Cavity      & 187.4  & 19.6    & 4.82 & 77.4 & 198.9 & 1229.8 &  9.56\\
\enddata
\tablecomments{The bolometric luminosity ($L$) of the source a--h are not derived due to the high contamination toward $F_{\rm tot}$ of warm temperature regime by background PDR emissions.}
\label{tb:virial}
\end{deluxetable*}

\subsection{Physical Properties of Extended Sources: Kinematic Status and Global History}\label{sec:es}

In \citetalias{2019ApJ...873...51L}, we studied the relative evolutionary states of molecular clumps in W\,51\,A by comparing their kinematic and physical properties by deriving and cross-correlating the virial parameter, $\alpha_{\rm vir}$, and the luminosity-to-mass ratio, $L/M$, of individual radio-defined extended sources. Comparison of the two independent clump evolutionary tracers, $\alpha_{\rm vir}$ and $L/M$, showed a correlation in log-space for W\,51\,A where lower $\alpha_{\rm vir}$ and $L/M$ indicated younger clump evolutionary states. In this section, we will apply the $\alpha_{\rm vir}$ versus $L/M$ analysis to the M\,17 molecular clumps in order to understand the star formation history of M\,17. We will first begin by discussing the techniques used.

\subsubsection{The Luminosity-to-mass Ratio and The Virial Analysis}\label{anal}

Here we briefly demonstrate how we derive the bolometric luminosity ($L$), mass ($M$), and virial parameter ($\alpha_{\rm vir}$). Since we follow the methods of \citetalias{2019ApJ...873...51L} with some necessary modifications, we will explain the general analyses as well as the difference between the methods used in this study and in the previous study.

We estimated the mass of each molecular clump from the mass surface density ($\Sigma$) map and the given distance (1.98~kpc). The $\Sigma$ map was derived by the pixel-by-pixel graybody fitting method that was investigated in \citet{2016ApJ...829L..19L}. We only used the \textit{Herschel}-PACS 160\,$\mu$m and  \textit{Herschel}-SPIRE 250, 350, and 500\,$\mu$m images for the analysis so we could assume the optically thin limit for the cold dust emission. We first convolved the four shorter wavelength \textit{Herschel} images to match to the angular resolution of the 500\,$\mu$m maps ($\sim$36$\arcsec$). We then applied a simple radiative transfer equation using the optically thin assumption (see Equation 1 in \citetalias{2019ApJ...873...51L}), and adopting the thin ice mantle dust opacity model of \citet{1994A&A...291..943O} and a dust-to-gas mass ratio of 1/142 \citep{2011ApJ...732..100D} to estimate dust opacity ($\kappa_{\nu}$). We utilized the temperature ($T$) maps at the 500\,$\mu$m resolution as a template to derive higher angular resolution $\Sigma$ maps. We repeated the graybody fit using the JCMT 850\,$\mu$m map with the template $T$ map (re-grided to match to the 850\,$\mu$m map). This enabled us to produce a high angular resolution $\Sigma$ map ($\sim$14$\arcsec$), which in turn allowed us to derive more accurate masses for the molecular clumps.

The bolometric luminosity ($L$) of each molecular clump defined in this study is derived from the two temperature graybody assumption as was performed in \citetalias{2019ApJ...873...51L}. The integrated intensity inside the defined aperture for each source at each image wavelength was calculated to perform the two temperature graybody fitting. For this we utilized the {\it Spitzer}-IRAC bands (3.6--8.0\,$\mu$m), \textit{SOFIA}-FORCAST 20 and 37\,$\mu$m, {\it Herschel}-PACS 70 and 160\,$\mu$m, {\it Herschel}-SPIRE 250, 350 and 500\,$\mu$m, and \textit{JCMT}-SCUBA2 850\,$\mu$m images. The JCMT 850\,$\mu$m data have sufficient resolution that we used their measured integrated intensities as nominal data points in this study. 
In addition to the photometric uncertainty levels of each band as described in Section \ref{sec:obs}, for the large extended regions defined in Figure \ref{fig:es} one needs to also consider the de-reddening effect, the contribution of different temperature components, and nearby source contamination as additional errors. Accounting for this in the same fashion as was done in \citetalias{2019ApJ...873...51L}, we assume $\sim$30\% total photometric uncertainty for 4.5\,$\mu$m, $\sim$40\% for 20 and 37\,$\mu$m, and $\sim$50\% for 70 and 160\,$\mu$m (Table \ref{tb:es}). Taking into account the possibility of PAH contamination, we treated the 3.6, 5.8, and 8\,$\micron$ intensities as upper limits. The 250, 350, 500, and 850\,$\micron$ bands were also considered as upper limits due to their poor angular resolution and the possibility of contamination by extended emission from the environment. We also de-reddened the flux of each wavelength by using the 1-D radiative transfer equation of absorption, $F_{\rm \nu,tot,1} \simeq F_{\rm \nu,tot,0}^{-\tau}$, where $F_{\rm \nu,tot,0}$ is the intrinsic flux and $F_{\rm \nu,tot,1}$ is the observed fluxes (i.e., with extinction). The opacity was estimated using the median $\Sigma$ of each clump so that $\tau_{\nu}$\,=\,1/2\,$\kappa_{\nu}$\,$\tilde{\Sigma}$.

We derived the virial parameter, $\alpha_{\rm vir}$, assuming constant density for the molecular clumps using Equation 2 in \citetalias{2019ApJ...873...51L}. The measurement of the gas kinematics needed for that equation (i.e., $\sigma$, the FWHM of the molecular line profile in km/s) was derived from the $^{13}$CO(1-0) data from the Nobeyama 45\,m telescope observations of M\,17 by \citet{2018PASJ...70S..42N} which have an angular resolution of $\sim$20$\arcsec$. The uncertainty of $\alpha_{\rm vir}$ was derived from the combined errors in the gas velocity width, derived clump mass, and distance estimation so that the conservative total uncertainty of $\alpha_{\rm vir}$ is estimated to be about a factor of 2 \citep[e.g.,][]{2013ApJ...779..185K}. 

\subsubsection{The Relative Evolutionary States of the Sub-Components of M~17}

\begin{figure}
\epsscale{1.1}
\plotone{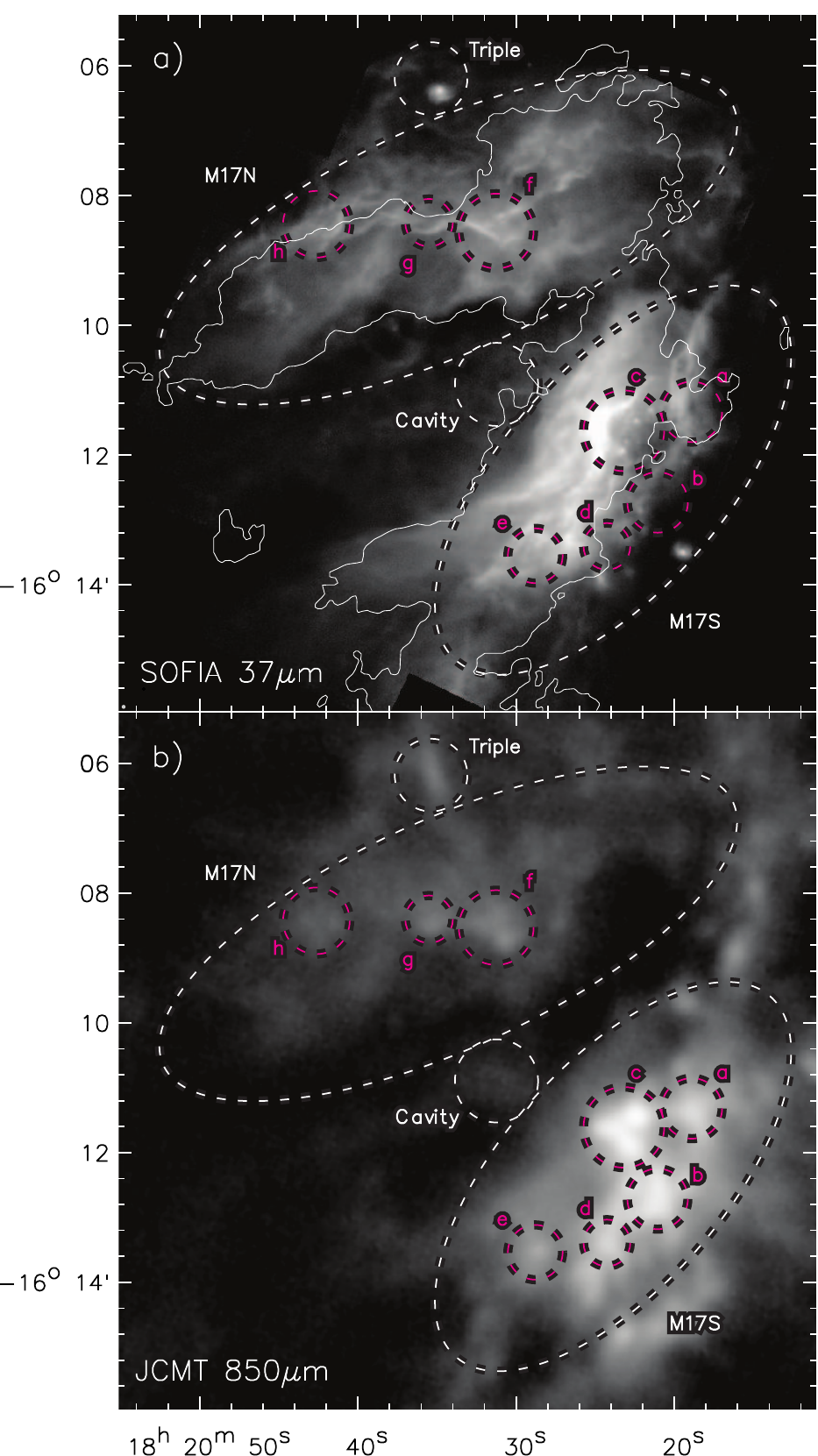}
\caption{\footnotesize The regions within M\,17 used in our investigation into the evolutionary states of the substructures within the G\ion{H}{2} region. a) The 37\,$\mu$m \textit{SOFIA} map, where the thin white lines are the lowest level contour of the 21\,cm radio continuum map of \citet{1984A&A...136...53F}. b) The \textit{JCMT} 850\,$\mu$m map of the same region. The large white ellipses in both panels are the sources for which there are analogous structures in the infrared, submillimeter, and radio continuum. The red circles are the submillimeter-defined peaks and substructures. These sources are referred to in Tables \ref{tb:es} and \ref{tb:virial}.}
\label{fig:es}
\end{figure}

\begin{figure}
\plotone{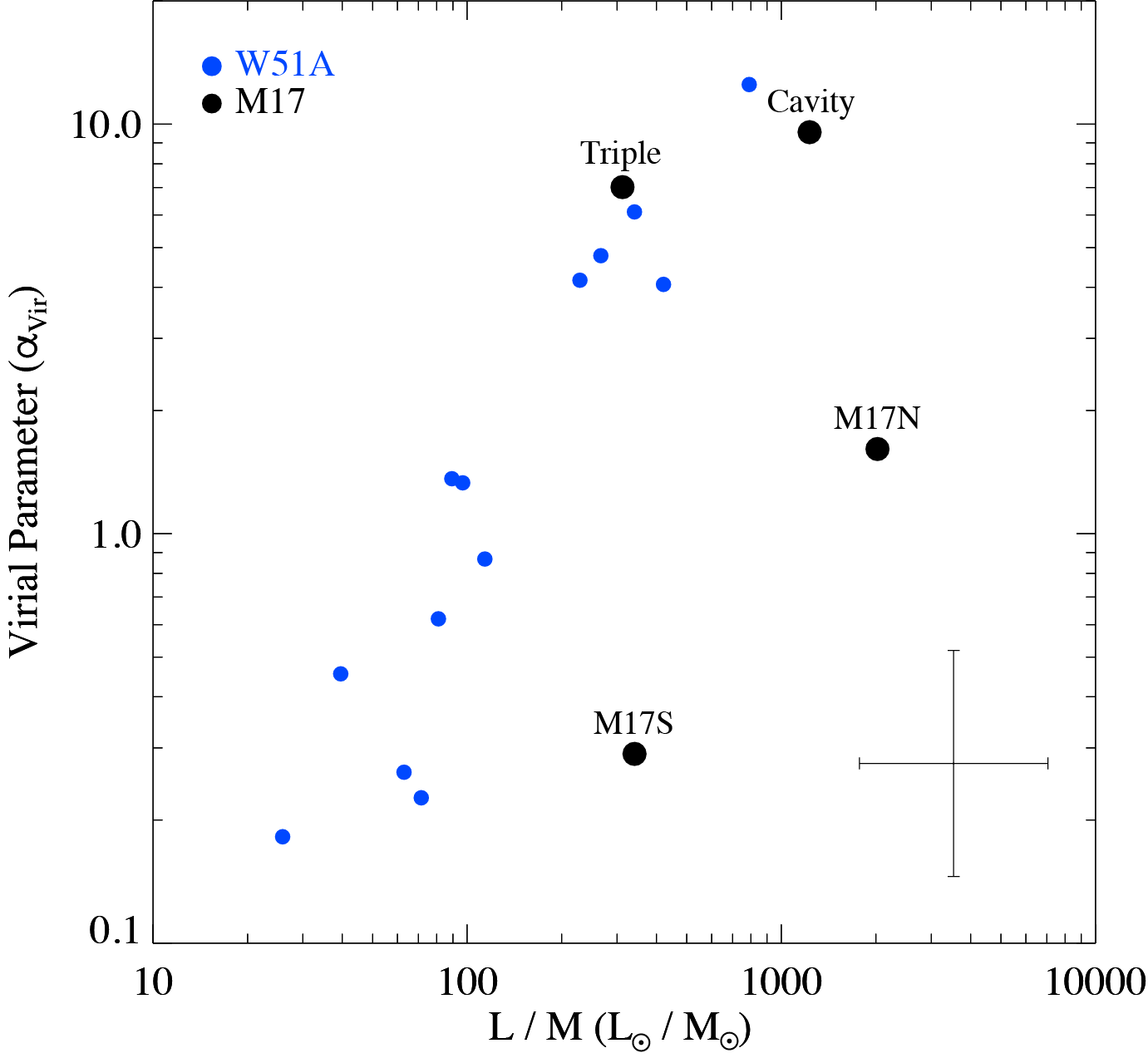}
\caption{\footnotesize The $\alpha_{\rm vir}$ vs. $L/M$ of mid-infrared extended sources which are potentially molecular clumps. The colored dots show individual clumps as indicated in the plot. The blue dots are the data points from W51A areas \citepalias{2019ApJ...873...51L}. The black dots are the molecular clump candidates of this study. The error bars drawn in the bottom-right corner represent the factor of 2 error for both $\alpha_{\rm vir}$ and $L/M$.}
\label{fig:alm}
\end{figure}

For W\,51\,A, we defined the extended but spatially separated radio sources that were already defined in previous studies \citep[e.g.,][]{1972MNRAS.157...31M} as the star-forming molecular clumps containing embedded massive young star clusters that are ionizing the extended \ion{H}{2} regions seen in radio continuum. To be consistent with that previous analysis, we look at the only two radio sub-regions of M\,17: M\,17\,N and M\,17\,S. As shown in Figure \ref{fig:es}, we create two ellipses that cover most of the radio emitting areas of M\,17\,N and M\,17\,S, based on the 21\,cm continuum map of \citet{1984A&A...136...53F} and give the sizes and derived mid-infrared fluxes within those ellipses in Table \ref{tb:es}. Using the techniques described in the previous subsection (\S\ref{anal}), we summarize the physical parameters we derived for both clumps in Table~\ref{tb:virial}, i.e., the virial mass ($M_{\rm vir}$), clump mass ($M$), bolometric luminosity ($L$), the derived warm and cold temperature components ($T_{\rm cold}$ and $T_{\rm warm}$), the luminosity-to-mass ratio ($L/M$), and the virial parameter ($\alpha_{\rm vir}$). We also placed the locations of M\,17\,N and M\,17\,S on the plot of virial parameter vs. luminosity-to-mass ratio in Figure \ref{fig:alm} along with the data from W\,51\,A. In this plot we see that M\,17\,N has a higher value of both $\alpha_{\rm vir}$ and $L/M$ than M\,17\,S, suggesting that M\,17\,N is more evolved. However, while these two points for M\,17 seem to show an agreement with the slope of the W\,51\,A data,  they are off by factor of $\sim$1.0\,dex in the direction of higher $L/M$. 
It is unknown if the slope of the $\alpha_{\rm vir}$ vs. $L/M$ relation has a universal value or not. As we continue to add data from other G\ion{H}{2} regions in this survey, we intend to explore this issue. If the slope is indeed universal, a reason for the offset of the M\,17\,N and M\,17\,S points from the relationship seen in W\,51\,A could be  that we are deriving much larger luminosity values for them due to the high levels of external heating of both M\,17 sub-regions by the revealed open cluster NGC\,6618. Such external heating contributes significantly to the bolometric luminosity in the warm temperature regime, i.e. $\lambda\sim$ 3.6--20$\micron$, and as we discussed in \citetalias{2019ApJ...873...51L}, the warm temperature component of the SED dominates the bolometric luminosity estimate (while the mass estimates of the clumps are more sensitive to the cold SED component).

While the evolutionary trend between M\,17\,N and M\,17\,S is believed to be real, one might also consider these regions to be too large for this type of analysis, i.e. using the whole of M\,17\,N and M\,17\,S to represent individual star-forming molecular clumps. The largest region in W\,51\,A that we considered to be a clump was G49.5-0.4b which measured 11.2\,pc$^2$, compared to 8.7\,pc$^2$ for M\,17\,N, though most were smaller. However, because the entire M\,17 G\ion{H}{2} region is pervaded at mid-infrared and radio continuum wavelengths by the heat and ionizing flux of the revealed cluster of NGC\,6618, we cannot easily distinguish in those maps any separated, smaller, star-forming molecular clumps that may exist.

In the far-infrared and submillimeter there is practically no contamination by either the heating or the ionizing flux of the NGC\,6618 cluster, so we turned there to look for smaller potential star-forming molecular clumps within M\,17. We examined the 850\,$\mu$m \textit{JCMT}-SCUBA map \citep{2006ApJ...644..990R} and selected the four brightest and most obvious 850\,$\mu$m clumps in M\,17\,N and five of the most obvious clumps in M\,17\,S (Figure\,\ref{fig:es}). These were found via an optimal extraction method that looked for structures on the scale of $\sim$20$\arcsec$ so as to match the size-scale of the sources seen in the (coarser resolution) CO data. One of the submillimeter clumps in the north corresponds to G015.128, or the Triple, and unlike the other submillimeter-defined clumps it does appear as a separate source in the mid-infrared from the bar of emission in M\,17\,N. In this analysis we name it `Triple'. We named the rest of the submillimeter regions from a to h in the order of ascending right ascension, with a--e belonging to M\,17\,S and f--h to M\,17\,N. We also decided to place an aperture around the central region of the NGC\,6618 cluster, which has diffuse mid-infrared and submillimeter emission, but does show substantial molecular line ($^{13}$CO) emissions, and we named the region `Cavity'. The locations and sizes of all of these regions are shown in Figure~\ref{fig:es}. Since both the Triple and Cavity are separate, distinguishable regions in the mid- and far-infrared, we added them to Table\,\ref{tb:es}, computed their physical parameters, including $\alpha_{\rm vir}$ and $L/M$ (see Table\,\ref{tb:virial}), and placed their locations in the plot of $\alpha_{\rm vir}$ vs. $L/M$ (Figure\,\ref{fig:alm}). Interestingly, the location of the Triple on the plot $\alpha_{\rm vir}$ vs. $L/M$ has it lying along the general trend seen for the W\,51\,A sources. If the slope of the trend in $\alpha_{\rm vir}$ vs. $L/M$ for W\,51\,A is universal, the Triple might lie along this trend because, unlike M\,17\,N and M\,17\,S, it is situated north of the extended radio emission of M\,17 and likely not influenced by the environmental heating of NGC\,6618. In any case, both evolutionary tracers point to the Triple as being a highly evolved clump, but not as evolved as the Cavity. The Cavity, as expected, is shown to have an extremely high virial parameter indicative of an expanding, unbound, clump which would be appropriate for the material left over in the area around such an evolved stellar cluster.

For the submillimeter-defined sources (a--h), we believe that their mid-infrared emission is likely dominated by external environmental heating and therefore estimates of the bolometric luminosity (and consequently their L/M ratio) cannot be trusted. We therefore calculated just their virial parameters as an evolutionary tracer since those calculations require far-infrared/submillimeter continuum and millimeter molecular line emission data only. These derived $\alpha_{\rm vir}$ values along with cold temperature fits to the long wavelength data and their derived masses are tabulated in Table\,\ref{tb:esvir2}. 

As we have stated, the overall conclusion from  Figure\,\ref{fig:alm} is that the M\,17\,S region appears relatively younger, while the source Cavity appeared as one the oldest regions in M\,17 then followed by Triple. This evolutionary trend is supported by the $\alpha_{\rm vir}$ values of source a--h. As shown in Table\,\ref{tb:esvir2}, the $\alpha_{\rm vir}$ of the clump candidates in M\,17\,S (a--e) are all under 2, meaning gravitationally bound and younger, while all of the sources in M\,17\,N (f--h) are all larger than 2, and thus unbound and older. 

\begin{deluxetable*}{rrrrrrrrr}
\centering
\tabletypesize{\scriptsize}
\tablecolumns{10}
\tablewidth{0pt}
\tablecaption{Observational and Physical Parameters of Submillimeter-Defined Clumps in M\,17}
\tablehead{ \colhead{  Source  }&
           \colhead{   R.A.  } &
           \colhead{   Dec.  } &
           \colhead{ $R_{\rm tot}$ } &
           \colhead{   $M_{\rm vir}$  } &
           \colhead{   $M$  } &
           \colhead{   $T_{\rm cold}$  } &
           \colhead{ $\alpha_{\rm vir}$ } \\
	   \colhead{  } &
	   \colhead{  } &
	   \colhead{  } &
	   \colhead{ ($\arcsec$) } &
	   \colhead{   ($M_{\sun}$)   } &
	   \colhead{   ($M_{\sun}$)   } &
	   \colhead{   ($K$)   } &
	   \colhead{             } }
\startdata
         a & 18 20 18.8 & -16 11 19.8 & 28.1 &  354.8 & 332.1  & 39.2 &  1.07 \\
         b & 18 20 20.9 & -16 12 42.8 & 28.1 &  276.4 & 582.1  & 37.6 &  0.47 \\
         c & 18 20 23.1 & -16 11 38.4 & 37.1 &  456.5 & 780.2  & 44.8 &  0.59 \\
         d & 18 20 24.1 & -16 13 24.4 & 21.2 &  185.0 & 216.1  & 41.8 &  0.86 \\
         e & 18 20 28.7 & -16 13 34.3 & 24.4 &  163.4 &  82.3  & 47.8 &  1.98 \\
         f & 18 20 31.1 & -16 08 32.2 & 34.7 &  151.7 &  52.0  & 45.8 &  2.92 \\
         g & 18 20 35.4 & -16 08 26.3 & 21.3 &   75.5 &  20.1  & 40.3 &  3.75 \\
         h & 18 20 42.5 & -16 08 27.7 & 30.9 &  178.0 &  14.4  & 44.9 & 12.36 \\
\enddata
\tablecomments{\footnotesize R.A. and Dec. are for the center of the apertures defined by \textit{$R_{\rm tot}$}. The $M_{\rm vir}$, $M$, $T_{\rm cold}$, and $\alpha_{\rm vir}$ are defined as same as Table\,\ref{tb:virial}}
\label{tb:esvir2}
\end{deluxetable*}

\subsubsection{The History of Stellar Cluster Formation in M17}\label{sec:hist}

\citet{2009ApJ...696.1278P} studied star formation history of the 1$\fdg 5\times$1$\degr$ area around M\,17. They claimed that an extended bubble ($\sim$30$\arcmin$ diameter, named M\,17\,EB) located to the north of M\,17 may have sequentially triggered star formation \citep{elmegreen1992} within the M\,17 cloud, including the formation of NGC\,6618 cluster. This conclusion was based on comparisons of the estimated YSO ages via YSO SED fitting \citep{2006ApJS..167..256R} toward the {\it Spitzer} point sources defined in \citet{2009ApJ...696.1278P} as well as the morphologies of the dense gas structures lining up around the rim of the M\,17\,EB expanding shell. The star cluster residing almost at the center of M\,17\,EB and the prominent high-mass stars inside M\,17\,EB show older YSO ages than the members of NGC\,6618. \citet{2008ApJ...686..310H} claimed the triggering star formation in M\,17 may have occurred only locally, i.e. by the expanding shell of the M\,17 \ion{H}{2} region into the northern and southern bars, instead of such a large scale triggering effect insisted by \citet{2009ApJ...696.1278P}.

Our evolutionary tracers, $\alpha_{\rm vir}$ and $L/M$, of the extended sources in M\,17 allow us to comment on the possibility of both external and local triggering scenarios. If triggering occurred only locally by the expansion of the ionization and shock fronts caused by the most massive stars at the center of NGC\,6618 into the molecular cloud that formed M\,17, then one would expect that the regions of M\,17\,N and M\,17\,S should have approximately the same evolutionary state, and therefore similar $\alpha_{\rm vir}$ and $L/M$ values, which is not the case. If M\,17 were created purely by external triggering by the expansion of M\,17\,EB from the north, we would expect a trend in our evolutionary tracers where regions would be older to the north and younger to the south (i.e., oldest to youngest: Triple, M\,17\,N, Cavity, M\,17\,S). What we see is that the Cavity seems to be the oldest region, inconsistent with this scenario. 

While we don't necessarily see the effects of triggering from M\,17\,EB, we might perhaps be seeing its influence kinematically on M\,17 in our derived virial parameters. Both the Triple and region h are located near the boundary of the expanding shell of M\,17\,EB and the expanding G\ion{H}{2} region of M\,17, and both have extremely high values of $\alpha_{\rm vir}$ (7.02 and 12.36, respectively). This may be due to the two shocks from M\,17\,EB and M\,17 \ion{H}{2} regions colliding around the location of source h, injecting a large level of kinetic energy there and resulting in what we see as the highest $\alpha_{\rm vir}$ value in the region.

\subsection{Present and future star formation in M~17}

\citet{1976ApJS...32..603L} claimed that, based on CO observations, the M\,17\,SW molecular cloud that peaks to the southwest of M\,17\,S should be dense enough to undergo collapse via self-gravitation. Consequently, if there is to be a likely area for present (and future) star formation activities, M\,17\,SW would be it. \citet{2008ApJ...686..310H} postulate that the KW Object, UC\,1, and IRS\,5 along with some other near-infrared excess sources likely represent a recent phase of star formation spatially independent of the revealed, large, OB cluster at the heart of the reflection nebula of M\,17. However, apart from these few prominent infrared-bright sources, the apparent dearth of identified YSOs has been pointed out by several authors \citep[e.g.,][]{2002ApJ...577..245J}. Our finding of only 7 sources in M\,17 that are likely to be MYSO candidates is indeed markedly smaller than the number of MYSO candidates (41) within W\,51\,A. However, our observed area of M\,17 is 5.8\,pc $\times$ 5.8\,pc, while for W\,51\,A we observed an area of approximately 27\,pc $\times$ 11\,pc (i.e. $\sim$9$\times$ larger). By a simple source-per-area argument and extrapolating from W\,51\,A, one would expect that M\,17 should house about 5 MYSO candidates, which is roughly consistent with the 7 MYSO candidates we catalogued. 

On the other hand, there does seem to be a dearth of YSOs when compared to the previous star formation episode that yielded the rich OB cluster of NGC\,6618. While that may be the case, M\,17 is, physically, a relatively small G\ion{H}{2} region and may not produce another episode of star formation like the previous one. The largest reservoir of material left in M\,17 is M\,17\,SW, which only contains $\sim$5$\times$10$^3$\,M$_{\sun}$ of gas \citep{2003ApJ...590..895W}, and therefore does not have enough mass to produce a large cluster of several hundred stars like NGC\,6618 (which has more than 100 stars more massive than spectral type B9) in the future. 

As for the present epoch of star formation, \citet{2016ApJ...833..193R} found that there are about as many X-ray protostars as infrared protostars in M\,17, many of which cannot be detected against the bright infrared nebular emission of M\,17. Because low-mass protostars demonstrate high levels of magnetic activity on their surfaces throughout their protostellar evolution, these X-ray sources are likely to be just that. It may be that the present epoch of star formation coming from the M\,17\,SW molecular cloud predominantly consists of low-mass YSOs. 

\section{Summary}\label{sec:sum}

In this, our second paper from our mid-infrared imaging survey of Milky Way Giant \ion{H}{2} regions, we obtained \textit{SOFIA}-FORCAST 20 and 37\,$\micron$ images toward the central $\sim$10$\arcmin\times$10$\arcmin$ area of M\,17. We compared these \textit{SOFIA}-FORCAST images with previous multi-wavelength observations from various ground- and space-based telescopes in order to inspect morphological and physical properties of compact and extended sources in M\,17. We itemize below the summary of major discoveries made in this study.

1) The significantly different appearance in morphology and brightness distribution of emission in M\,17 at 20\,$\micron$ when compared to any other infrared wavelength implies that it traces different physics within the same environment. Spectra of regions within M\,17 show that the brightness of the [\ion{S}{3}] line at 18.71\,$\micron$ tracks with the brightness of the emission seen in the 20\,$\micron$ filter of FORCAST (whose bandpass is $\sim$17 to $\sim$23\,$\mu$m). The 20\,$\mu$m image, therefore, seems to trace better the large-scale ionized gas structure of M\,17, while the 37\,$\micron$ image traces the dust continuum emission, making its appearance more consistent with {\it Spitzer}-IRAC and \textit{Herschel}-PACS images.

2) Previous near-infrared images resolved the Kleinmann-Wright object into a binary \citep{2004aap...427..849C}, however we determine that only KW-1 appears to have any significant emission in the mid-infrared. The object is seen as a source separate from the M\,17\,S infrared-emitting region, and lying more than 0.6\,pc from any other mid-infrared point source.  This may indicate KW is an isolated system, and not, as previously suggested, a central member of a young stellar cluster \citep{2004aap...427..849C}.  We determine that KW has a mass of 8\,M$_{\sun}$, in keeping with previous suggestions that it is perhaps a Herbig Be object \citep{2004aap...427..849C}. We resolve the emission from the KW object and find that it is extended at 20 and 37\,$\micron$. This extension is consistent with the angle of the bipolar polarization pattern seen by \citet{2012PASJ...64..110C}, suggesting that the mid-infrared morphology may be influenced by an outflow from within the KW object.

3) The two brightest infrared sources in the M\,17\,S region are UC\,1 and IRS\,5. While IRS\,5 has no detected radio continuum emission, UC\,1 is a bright hypercompact \ion{H}{2} region \citep[e.g.,][]{2012ApJ...755..152R}. Our modeling of the SED of UC\,1 suggests that it is the most massive MYSO in M\,17, weighing in at 64\,M$_{\sun}$. While UC\,1 and IRS\,5 have comparable brightnesses at wavelengths $<$20\,$\mu$m, IRS\,5 is significantly fainter at 37\,$\mu$m and not seen in the Herschel 70\,$\mu$m images. It cannot be well-fit by our MYSO model fitter. We suggest it is a intermediate-mass Class II object and not a MYSO. 

4) Previous near-infrared studies of G015.128 showed it to be a triple source \citep{2002ApJ...577..245J, 2012PASJ...64..110C}, however we find that only one of these sources, T1, dominates the emission at 20 and 37\,$\micron$.

5) In addition to seven previously identified infrared sources, we detect and identify an additional nine new compact sources in M\,17 at 20 and 37\,$\mu$m. However, we do not detect most previously identified YSOs and/or high-mass Class I sources identified at near-infrared wavelengths. This is likely due to the fact that some sources have been misclassified, or are lower-mass and/or more evolved sources and not true MYSOs.

6) We utilized \textit{Spitzer}-IRAC, \textit{SOFIA}-FORCAST, and \textit{Herschel}-PACS photometry data to construct SEDs of all 16 compact sources identified. We fit the SEDs with MYSO models and found 7 sources that are candidate MYSOs based on those fits. This is significantly fewer MYSOs than identified in the W\,51\,A G\ion{H}{2} region in \citetalias{2019ApJ...873...51L}. We suggest that differences in size and distance of the two G\ion{H}{2} regions are likely at play.   

7) We calculated the luminosity-to-mass ratio and virial parameters of the extended sub-regions of M\,17 to estimate their relative ages quantitatively. The results suggest that M\,17\,S is younger than M\,17\,N. This seems consistent with the fact that the bulk of the dense molecular material in the region exists in the M\,17\,S/M\,17\,SW area, and that this is also where we find the vast majority of the presently forming YSOs we detect at 20 and 37\,$\mu$m.

\acknowledgments
Authors thank an anonymous referee for constructive comments that help to improve the manuscript significantly. Authors also thank B.-G. Andersson, J. T. Schmelz, J. C. Tan, W. D. Vacca and Y. Zhang for discussions. This research is based on observations made with the NASA/DLR Stratospheric Observatory for Infrared Astronomy (\textit{SOFIA}). \textit{SOFIA} is jointly operated by the Universities Space Research Association, Inc. (USRA), under NASA contract NAS2-97001, and the Deutsches \textit{SOFIA} Institut (DSI) under DLR contract 50 OK 0901 to the University of Stuttgart. This work is also based in part on archival data obtained with the Spitzer Space Telescope, which is operated by the Jet Propulsion Laboratory, California Institute of Technology under a contract with NASA. This work is also based in part on archival data obtained with Herschel, an European Space Agency (ESA) space observatory with science instruments provided by European-led Principal Investigator consortia and with important participation from NASA. This work is also based in part on observations with ISO, an ESA project with instruments funded by ESA Member States (especially the PI countries: France, Germany, the Netherlands and the United Kingdom) and with the participation of ISAS and NASA. Financial support for this work was provided by NASA through \textit{SOFIA} awards 03\_0008, 03\_0009, 04\_0001, 04\_0002, and 05\_0008 issued by USRA. 

\vspace{5mm}
\facility{\textit{SOFIA}(FORCAST)}

\clearpage

\appendix
\section{Data release}

The fits images used in this study are publicly available at: {\it https://dataverse.harvard.edu/dataverse/SOFIA-GHII}. 

The data include the \textit{SOFIA} FORCAST 20 and 37\,$\mu$m final image mosaics and their exposure maps, as well as the individual 20 and 37\,$\mu$m images of Source 1.

\section{Discussion of Interesting Areas with M~17}\label{appendixb}
Apart from the individual sources, there are a couple of regions of M\,17 that are particularly interesting as seen in the 20 and 37\,$\mu$m data, especially when comparing them to other wavelengths. Both regions contain ``pillars'' that reside within the central cavity of M\,17, but have unique differences. 

\begin{figure*}[htb!]
\epsscale{1.15}
\plotone{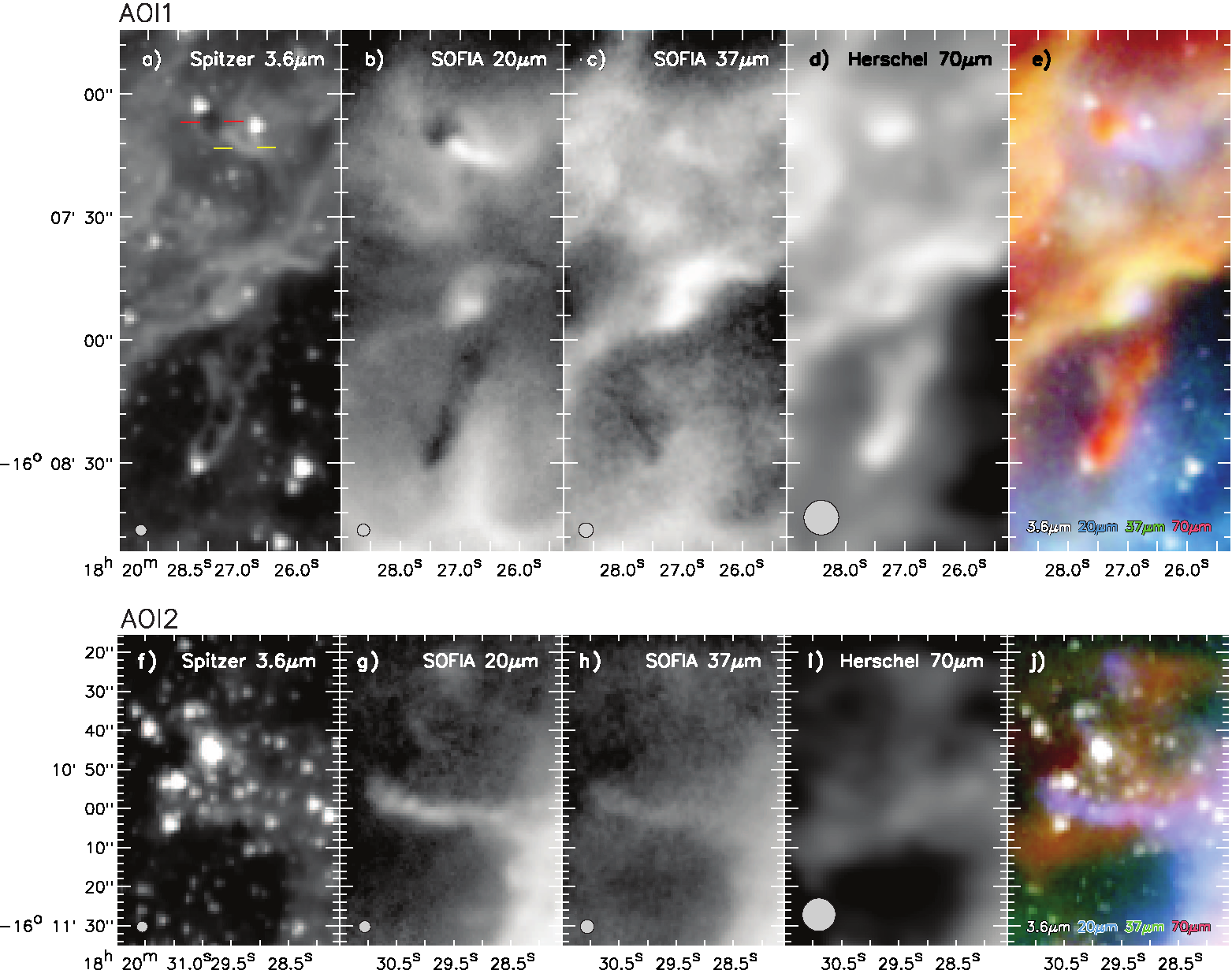}
\caption{Area of Interest 1 (a-e) and Area of Interest 2 (f-j). The wavelength is given in the upper right and the resolution of the images are shown at the lower left corner of each panel that shows a single wavelength image. On the bottom of the RGB image panelss the wavelengths representing each color are given. The red lines in a) lie to either side of the negative source and the yellow lines lie to either side of the bright arc discussed in the text. \label{fig:aoiboth}}
\end{figure*}

\subsection{Area of Interest 1}
Located on the inner cavity wall of M\,17\,N there are a few interesting structures separated by about an arcminute in declination. The first is a ``pillar'' of dust, seen as a dark area against the bright nebular background in the \textit{SOFIA} 20\,$\mu$m (Figure \ref{fig:aoiboth}b). In the \textit{Spitzer}-IRAC 3.6\,$\mu$m image, the dark pillar is outlined in emission (Figure \ref{fig:aoiboth}a). Because this emission is not seen at longer wavelengths, this is likely to be either scattered light off of the surface of the pillar and/or the hot dust emission on the outermost surface of the pillar. Looking to the \textit{Herschel} data at 70\,$\mu$m (Figure \ref{fig:aoiboth}d), one can see that the pillar is now seen brightly in emission. At 37\,$\mu$m  the source can not be easily differentiated from the background emission (Figure \ref{fig:aoiboth}c). Given this behavior as a function of wavelength it can be ascertained that the pillar is comprised of cold and dense material and is located in the foreground part of the nebula. The pillar also protrudes from the wall of M\,17\,N and points towards the central members of the ionizing cluster NGC\,6618. At 3.6\,$\mu$m, there is also a star seen at the tip of this pillar. It is therefore likely that the dense core that formed this star shielded the pillar from photo-evaporative erosion caused by the brightest stars of NGC\,6618, and is thus the reason for the pillar's existence. 

About an arcminute to the north of the pillar is another source that behaves similarly to the pillar as a function of wavelength, yet is a compact source (it lies between the red lines in Figure \ref{fig:aoiboth}a). Like the pillar, at 20\,$\mu$m the source is a negative. It is also seen as a negative source at 3.6\,$\mu$m, however it emits quite brightly at 70\,$\mu$m. Also like the pillar, at 37\,$\mu$m the negative source can not be easily differentiated from the background emission. This is therefore likely to be a colder molecular core located in the foreground part of the nebula.

There is also a very bright arc-like structure, seeming to radiate away from the negative source pointing to the southwest (it lies between the yellow lines in Figure \ref{fig:aoiboth}a). This arc is the brightest structure in the area at 20\,$\mu$m, yet is only modestly bright at 3.6 and 37\,$\mu$m (and not detected at 70\,$\mu$m). Though they are close in proximity, the negative source and the arc are likely not related. North of the bright arc there is a star seen at 3.6\,$\mu$m and it appears the arc is wrapping partially around it. Therefore, it is possible that the arc is being heated by this star that is seen at 3.6\,$\mu$m. However, the strong 20\,$\mu$m emission of the arc is likely due to ionization which can generate substantial [\ion{S}{3}] emission at 18.71\,$\mu$m, enhancing the flux observed in that filter (Appendix \ref{appendixd}). It is unlikely that the nearby 3.6\,$\mu$m star is responsible for this ionization, since no compact radio continuum emission source was detected at this location by \citet{2012ApJ...755..152R}. It is therefore more likely that the arc is being ionized and heated from the south by the more massive stellar members in the center of the NGC\,6618 cluster.

\subsection{Area of Interest 2}
In the center of the observed field, there is another mid-infrared source that appears to look like a ``pillar'' within the inner cavity of M\,17, this time protruding from the inner wall of M\,17\,S (see Figures 2 and 3). It is narrow in width ($\le$8$\arcsec$) and about 40$\arcsec$ in length (0.4\,pc). However, it behaves very differently as a function of wavelength than the previously mentioned pillar (Figure \ref{fig:aoiboth} a-e). Rather than being a negative source at 20\,$\mu$m, it is brightest at that wavelength (Figure \ref{fig:aoiboth} g). It is barely visible at 3.6 and 37\,$\mu$m (Figure \ref{fig:aoiboth} f\&h), and only the very tip is visible at 70\,$\mu$m (Figure \ref{fig:aoiboth} i). Given this behavior as a function of wavelength, it is like not a true pillar of material. At 3.6\,$\mu$m (Figure \ref{fig:aoiboth} f) it appears that the majority of revealed stars are north of this feature, and at 70\,$\mu$m there is a dense (and hence brighter) area of cold dust to the south of the feature. Therefore, the morphology at 20\,$\mu$m might be an edge-on view of the interface between a ridge of dust and the ionizing and heating stars interior to it.

\begin{deluxetable}{rrrrrrrrrr}
\tabletypesize{\scriptsize}
\tablecolumns{10}
\tablewidth{0pt}
\tablecaption{The Literature-Based Observational Parameters of the KW Object}
\tablehead{\colhead{   }&
           \colhead{ ${3.92\micron}$ } &
           \colhead{ ${4.64\micron}$ } &
           \colhead{ ${8.28\micron}$ } &
           \colhead{ ${8.70\micron}$ } &
           \colhead{ ${12.13\micron}$ } &
           \colhead{ ${14.65\micron}$ } &
           \colhead{ ${17.72\micron}$ } &
           \colhead{ ${21.34\micron}$ } \\
}
\startdata
         $F_{\rm int-bg}$ (Jy) & 4.40 & 5.77 & 22.8 & 30.9 & 32.6 & 33.1 & 55.5 & 68.8  \\
         Uncertainty (\%) & 15 & 15 & 20 & 10 & 20 & 20 & 25 & 30  \\
\enddata
\tablecomments{\footnotesize The mid-infrared flux photometry of KW source. The flux values are adopted from \citet{2004aap...427..849C}.}
\label{tb:kw}
\end{deluxetable}
 
\begin{deluxetable}{rrrrrrrr}
\tabletypesize{\scriptsize}
\tablecolumns{9}
\tablewidth{0pt}
\tablecaption{The Literature-Based Observational Parameters of the CEN\,92, IRS\,5, and UC\,1 Objects}
\tablehead{\colhead{ Sources }&
           \colhead{ $F_{8.7\micron}$\tablenotemark{a}} &
           \colhead{ $F_{9.8\micron}$\tablenotemark{b} } &
           \colhead{ $F_{10.38\micron}$\tablenotemark{a} } &
           \colhead{ $F_{10.53\micron}$\tablenotemark{b} } &
           \colhead{ $F_{11.7\micron}$\tablenotemark{b} } &
           \colhead{ $F_{11.85\micron}$\tablenotemark{c} } &
           \colhead{ $F_{17.72\micron}$\tablenotemark{a} } \\
           \colhead{  }&
           \colhead{ (Jy) } &
           \colhead{ (Jy) } &
           \colhead{ (Jy) } &
           \colhead{ (Jy) } &
           \colhead{ (Jy) } &
           \colhead{ (Jy) } &
           \colhead{ (Jy) } \\
}
\startdata
         CEN\,92 & \nodata & $1.5\pm0.2$ & \nodata & $1.8\pm0.3$ & $2.1\pm0.1$ & \nodata & \nodata \\
         UC\,1 & $18.7\pm1.3$ & $6.8\pm0.4$ & $7.3\pm1.0$ & $9.4\pm0.6$ & $25.4\pm1.3$ & $31.3\pm1.1$ & $114.67\pm29.7$ \\
         IRS\,5 & $3.2\pm1.6$ & $7.0\pm0.4$ & $6.8\pm1.5$ & $8.0\pm0.5$ & $10.6\pm0.6$ & $9.7\pm1.1$ & $130.0\pm31.0$ \\
\enddata
\tablecomments{\footnotesize The mid-infrared flux photometry of CEN\,92, UC\,1, and IRS\,5 sources. The flux values are adopted from \citet{2002AJ....124.1636K} and \citet{2015AAp..578A..82C}. The fluxes for CEN\,92 are only from \citet{2002AJ....124.1636K}. }
\tablenotetext{a}{The TIMMI2 bands on 3.6\,m ESO telescope \citep{2015AAp..578A..82C}.}
\tablenotetext{b}{The MIRAC2 bands on IRTF \citep{2002AJ....124.1636K}.}
\tablenotetext{c}{The VISIR band on VLT \citep{2015AAp..578A..82C}.}
\label{tb:others}
\end{deluxetable}

\begin{deluxetable}{lrrrrrrrrr}
\tabletypesize{\scriptsize}
\tablecolumns{8}
\tablewidth{0pt}
\tablecaption{{\it Spitzer}-IRAC bands Observational Parameters of Compact Sources in M\,17}
\tablehead{\colhead{  }&
           \colhead{  }&
           \multicolumn{2}{c}{${\rm 3.6\mu{m}}$}&
           \multicolumn{2}{c}{${\rm 4.5\mu{m}}$}&
           \multicolumn{2}{c}{${\rm 5.8\mu{m}}$}&
           \multicolumn{2}{c}{${\rm 8.0\mu{m}}$}\\
           \cmidrule(lr){3-4} \cmidrule(lr){5-6} \cmidrule(lr){7-8} \cmidrule(lr){9-10}\\
           \colhead{ Source }&
           \colhead{ $R_{\rm int}$ } &
           \colhead{ $F_{\rm int}$ } &
           \colhead{ $F_{\rm int-bg}$ } &
           \colhead{ $F_{\rm int}$ } &
           \colhead{ $F_{\rm int-bg}$ } &
           \colhead{ $F_{\rm int}$ } &
           \colhead{ $F_{\rm int-bg}$ } &
           \colhead{ $F_{\rm int}$ } &
           \colhead{ $F_{\rm int-bg}$ } \\
	   \colhead{  } &
	   \colhead{ ($\arcsec$) } &
	   \colhead{ (Jy) } &
	   \colhead{ (Jy) } &
	   \colhead{ (Jy) } &
	   \colhead{ (Jy) } &
	   \colhead{ (Jy) } &
	   \colhead{ (Jy) } &
	   \colhead{ (Jy) } &
	   \colhead{ (Jy) } \\
}
\startdata
       IRS5\tablenotemark{a} &  4.2 &   0.4610  &   0.3343  &  0.5997  &   0.3866   &   2.7970   &  1.8334    &  \nodata    & \nodata \\
        UC1\tablenotemark{a} &  3.0 &   0.6207  &   0.4130  &  1.4319  &   1.1007   &   6.4691   &  5.0239    &  \nodata    & \nodata \\
      CEN92\tablenotemark{a} &  3.0 &   0.3210  &   0.2301  &  0.3377  &   0.2427   &   0.9348   &  0.3825    &   2.6506    & 0.8838  \\
      Anon1                  &  3.0 &   0.0528  &   0.0039  &  0.0812  &   0.0148   &   0.3844   &  0.0577    &   1.3252    & 0.0614  \\
      Anon3\tablenotemark{a} &  4.2 &   0.3299  &   0.0851  &  0.3680  &   0.1012   &   2.5208   &  0.9260    &   7.8844    & 2.4335  \\  
   G015.128\tablenotemark{a} & 10.2 &   0.1004  &   0.0453  &  0.0992  &   0.0443   &   0.6705   &  0.1092    &   1.6035    & 0.1930  \\
          1\tablenotemark{a} &  5.4 &   0.2547  &   0.1818  &  0.3073  &   0.2439   &   3.9209   &  2.4356    &  11.1693    & 6.4071  \\
          2\tablenotemark{a} &  3.0 &   0.0308  &   0.0023  &  0.0328  &   0.0049   &   0.3697   &  0.0349    &   0.9896    & 0.0543  \\
          3\tablenotemark{a} &  3.0 &   0.0410  &   0.0064  &  0.0425  &   0.0098   &   0.4488   &  0.0841    &   1.2320    & 0.0811  \\
          4                  &  2.4 &   0.2602  &   0.1911  &  0.4978  &   0.4272   &   1.3501   &  0.8186    &   2.1925    & 0.6750  \\
          5                  &  3.0 &   0.1065  &   0.0346  &  0.2629  &   0.1739   &   0.7691   &  0.3349    &   2.0103    & 0.4474  \\
          6\tablenotemark{a} &  4.8 &   0.2005  &   0.0692  &  0.1973  &   0.0597   &   1.5262   &  0.2568    &   4.3840    & 0.5995  \\
          7\tablenotemark{a} &  9.0 &   0.8138  &   0.3330  &  0.8743  &   0.3941   &   5.4793   &  1.8169    &  16.4378    & 4.8067  \\
          8\tablenotemark{a} &  3.0 &   0.4669  &   0.4234  &  0.9658  &   0.9100   &   2.6997   &  2.3341    &  \nodata    & \nodata \\
          9\tablenotemark{a} &  2.4 &   0.0575  &   0.0193  &  0.0464  &   0.0113   &   0.2962   &  0.0580    &   1.0350    & 0.1390  \\
%         20 &  2.4 &   0.4223  &   0.3796  &  0.4486  &   0.4085   &   0.8340   &  0.5414    &   1.1397    & 0.3604  \\
\enddata
\tablecomments{\footnotesize Same as Table\,\ref{tb:cps1} but for \textit{Spitzer}-IRAC bands. The center positions of the apertures are based on \textit{SOFIA} observation in Table\,\ref{tb:cps1}. The KW source is not included in this table due to the saturation at all {\it Spitzer}-IRAC bands. See Table~\ref{tb:kw} for the flux photometry of KW source. Sources with no data at 8\,$\mu$m are saturated in that band.}
\tablenotetext{a}{PAH-contaminated sources as determined by the color-color diagram analysis in Fig.\,\ref{fig:ccd}.}
\label{tb:cps2}
\end{deluxetable}
 
\begin{deluxetable}{rrlrl}
\tabletypesize{\scriptsize}
\tablecolumns{8}
\tablewidth{0pt}
\tablecaption{{\it Herschel}-PACS bands Observational Parameters of Compact Sources in M\,17}
\tablehead{\colhead{  }&
           \multicolumn{2}{c}{${\rm 70\mu{m}}$}&
           \multicolumn{2}{c}{${\rm 160\mu{m}}$}\\
           \cmidrule(lr){2-3} \cmidrule(lr){4-5}\\
           \colhead{ Source }&
           \colhead{ $R_{\rm int}$ } &
           \colhead{ $F_{\rm int}$ } &
           \colhead{ $R_{\rm int}$ } &
           \colhead{ $F_{\rm int}$ } \\
	   \colhead{  } &
	   \colhead{ ($\arcsec$) } &
	   \colhead{ ($\times10^3$Jy) } &
	   \colhead{ ($\arcsec$) } &
	   \colhead{ ($\times10^3$Jy) } \\
}
\startdata
         KW & 16.0 &  0.55 & 22.5 &  1.73\tablenotemark{u}  \\
       IRS5 & 16.0 &  5.91\tablenotemark{u} & 22.5 &  4.85\tablenotemark{u}  \\
        UC1 & 16.0 &  2.82 & 22.5 &  4.26\tablenotemark{u}  \\
      CEN92 & 16.0 &  3.43\tablenotemark{u} & 22.5 &  4.37\tablenotemark{u}  \\
      Anon1 & 16.0 &  3.93\tablenotemark{u} & 22.5 &  5.79\tablenotemark{u}  \\
      Anon3 & 16.0 &  3.89\tablenotemark{u} & 22.5 &  5.27\tablenotemark{u}  \\
   G015.128 & 16.0 &  0.43 & 22.5 &  0.94\tablenotemark{u}  \\
          1 & 16.0 &  0.12 & 22.5 &  0.55\tablenotemark{u}  \\
          2 & 16.0 &  2.15\tablenotemark{u} & 22.5 &  3.34\tablenotemark{u}  \\
          3 & 16.0 &  2.16\tablenotemark{u} & 22.5 &  3.51\tablenotemark{u}  \\
          4 & 16.0 &  2.39\tablenotemark{u} & 22.5 &  2.30\tablenotemark{u}  \\
          5 & 16.0 &  3.98\tablenotemark{u} & 22.5 &  5.58\tablenotemark{u}  \\
          6 & 16.0 &  1.04\tablenotemark{u} & 22.5 &  1.97\tablenotemark{u}  \\
          7 & 16.0 &  1.77\tablenotemark{u} & 22.5 &  2.85\tablenotemark{u}  \\
          8 & 16.0 &  0.61\tablenotemark{u} & 22.5 &  0.69\tablenotemark{u}  \\
%         20 & 16.0 &  0.62\tablenotemark{u} & 22.5 &  0.83\tablenotemark{u}  \\
          9 & 12.8 &  0.03 & 13.5 &  0.03  \\
\enddata
\tablenotetext{u}{The $F_{\rm int}$ value is used as the upper limit since the source is not well resolved in the band.}
\tablecomments{\footnotesize Same as Table\,\ref{tb:cps2} but for \textit{Herschel}-PACS 70 and 160\,$\mu$m observation.}
\label{tb:cps3}
\end{deluxetable}

\section{Additional Photometry of Compact Sources in M\,17}\label{appendixc}

As discussed in \S\,\ref{sec:data}, in addition to the fluxes derived from the \textit{SOFIA}-FORCAST data, we used some additional photometry data in or SED analyses from the literature, as well as measured fluxes for our sources from both \textit{Spitzer}-IRAC and \textit{Herschel}-PACS. For the KW object we adopted the additional fluxes tabulated in Table\,\ref{tb:kw}. For CEN\,92, IRS\,5, and UC\,1, we adopted the additional fluxes tabulated in Table\,\ref{tb:others}. 

As we mentioned in \S\,\ref{sec:cps}, we performed optimal extraction photometry for the FORCAST 20 and 37\,$\mu$m images to define the location of all compact sources, and to determine the aperture radii to be used for photometry. Using these source locations, we employed the optimal extraction technique on the \textit{Spitzer}-IRAC 8\,$\mu$m data for all sources to find the optimal aperture to be used for all IRAC bands (since the source sizes are typically similar or smaller at the shorter IRAC bands). As we have done for the FORCAST images, we estimated the background emission from the annuli that showed the least contamination from nearby sources, i.e. showing relatively flat radial intensity profile (\S\,\ref{sec:cps}). Table\,\ref{tb:cps2} shows the photometry values we derive for all sources from the \textit{Spitzer}-IRAC bands.

Table\,\ref{tb:cps3} shows the photometry result for the \textit{Herschel}-PACS bands. We use fixed aperture radii for both PACS bands ($R_{\rm int}$=16$\farcs$0 for 70\,$\mu$m and $R_{\rm int}$=22$\farcs$5 for 160\,$\mu$m), except for source\,9 whose aperture is based on the PSFs of relatively isolated sources. In general, this aperture size cannot be determined accurately using the optimal extraction technique due to the ubiquity of extended emission from nearby sources that are overlapping the source being measured. We compared our aperture sizes to those typically used in the Hi-GAL Compact Source Catalogue \citep{2017MNRAS.471..100E,2016A&A...591A.149M}. That catalogue employs aperture sizes comparable to the ones we used in this study. We therefore believe that the fixed aperture sizes we employ here are reasonable, especially since the data are only being used to provide upper limits to our SED model fits in most cases. 

\begin{figure*}[htb!]
\epsscale{1.15}
\plotone{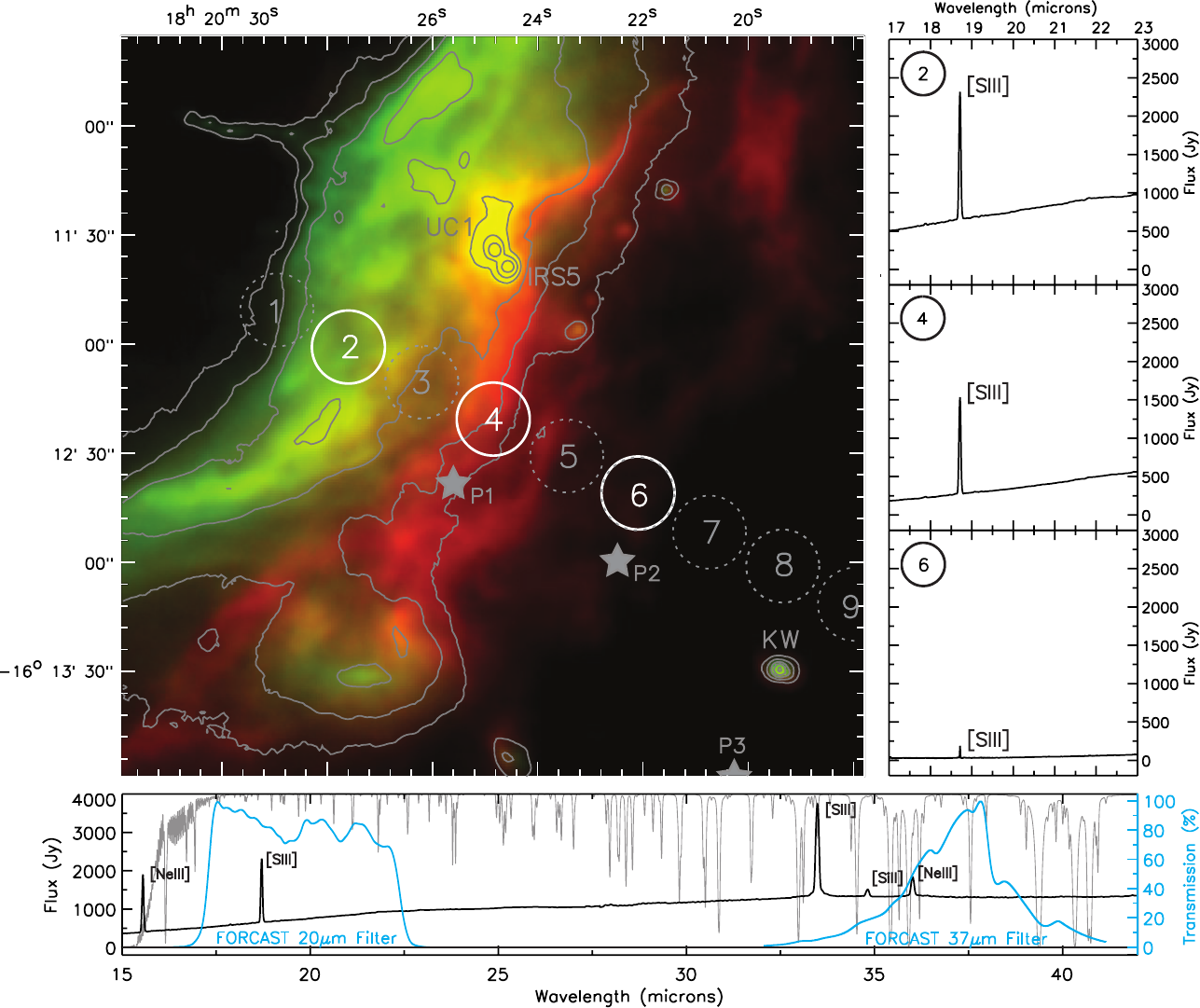}
\caption{The top left is a two-color image of a region of M\,17\,S where green is \textit{SOFIA}-FORCAST 20\,$\mu$m and red is \textit{SOFIA}-FORCAST 37\,$\mu$m. For additional clarity we overlay gray contours from the FORCAST 20\,$\mu$m image. The gray stars are the locations of spectral observations P1-P3 from \citet{2007ApJ...660..346P}. The large numbered circles show the positions of the spectra taken by \text{ISO}, and the size of the circles approximates the equivalent area that is being sampled by the \text{ISO} spectrometer at 20\,$\mu$m. The large circles displayed in white correspond to the locations of the \text{ISO} spectra which are shown in the right column of plots. The three spectra in the right column are shown with a wavelength range equal to that of the FORCAST 20\,$\mu$m passband, and it appears these contain only dust continuum and [\ion{S}{3}] line emission at all positions. The panel along the bottom of the figure shows a spectrum from 15 to 42\,$\mu$m at the location of position 2. Overplotted in blue are the FORCAST 20\,$\mu$m and 37\,$\mu$m filter profiles normalized so that their peak transmissions are at 100\%. Overplotted in gray is the atmospheric transmission of a typical SOFIA observation (i.e. 41,000\,ft aircraft altitude, telescope at a zenith angle of 45$\arcdeg$, with 7\,$\mu$m of precipitable water vapor overburden) using the ATRAN model \citep{Lord92}. All \textit{ISO} spectra shown are fully reprocessed spectra from \citet{2003ApJS..147..379S}. \label{fig:app}}
\end{figure*}

\section{Contamination by [\ion{S}{3}] in the FORCAST 20 micron Filter in Ionized Regions}\label{appendixd}

As discussed in \S\,\ref{sec:results1}, the appearance of the large-scale emission of M\,17 looks different in the images taken with the FORCAST 20\,$\mu$m filter than the images at any other wavelength we used in this study. In fact the shorter wavelength images , i.e. \textit{Spitzer}-IRAC 3.6, 4.5, and 5.8\,$\mu$m images (8.0\,$\mu$m is saturated), look more similar to the longer wavelength images, i.e. FORCAST 37\,$\mu$m and \textit{Herschel}-PACS 70\,$\mu$m, than what is seen in the FORCAST 20\,$\mu$m images. The only possible explanation is that there is some other form of emission other than dust continuum emission present which emits brightly at wavelengths between $\sim$17 and $\sim$23\,$\mu$m (i.e. the FORCAST 20\,$\mu$m bandpass). 

\citet{2007ApJ...660..346P} showed \textit{Spitzer}-IRS 9.9-19.6\,$\mu$m spectra (with slit size 4$\farcs$7$\times$11$\farcs$3 and a spectral resolving power of $\sim$600) taken at four discrete locations in the M\,17\,SW region which were labeled P1-P4, where the distance from the center of M\,17 increased from P1 to P4. There is an emission line seen at 18.71\,$\mu$m from [\ion{S}{3}] at each of these locations, and there is a trend where the [\ion{S}{3}] line was brightest at P1 ($\sim$240 Jy) and decreased in brightness from P1 to P4. [\ion{S}{3}] is a known tracer of ionized gas in \ion{H}{2} regions (e.g. \citealt{2006ApJ...639..788D}), so the drop off in line brightness as one moves away from the brightest regions of free-free emission was to be expected. Three of the locations from \citet{2007ApJ...660..346P} are shown in Figure \ref{fig:app} (the fourth lies further off the image to the southwest). Only P1 was in an area where we detect significant 20\,$\mu$m emission with FORCAST. 

We found additional unpublished data in the \textit{ISO} archive, taken at positions that corresponded better to the location of the 20\,$\mu$m emission. These were part of a program where ten 2.38--45.21\,$\mu$m spectra (i.e. the SWS01 mode, with spectral resolving power of 1000-2500) were taken at positions stepped perpendicularly across the entire M\,17\,S bar (Figure \ref{fig:app}). In these spectra we see that there is a definite trend of brighter [\ion{S}{3}] line emission with brighter 20\,$\mu$m flux that holds across all ten of the spectra sampled. We chose three representative positions to demonstrate this and show them in Figure  \ref{fig:app}. If one were to take a cross-sectional cut through our 20\,$\mu$m data along the series of \textit{ISO} positions, the location of position 2 (ISO Observation ID TDT09900212) would be near the maximum 20\,$\mu$m flux; position 4 (TDT09900214) would be near the drop-off in our detected 20\,$\mu$m emission, and position 6 (TDT32900866) would be in an area where we see very little emission at 20\,$\mu$m. The spectra from these three positions are also shown in right-hand panels of Figure \ref{fig:app}, and it can be seen that the [\ion{S}{3}] line emission brightness at 18.71\,$\mu$m trends with the FORCAST 20\,$\mu$m flux, with the brightest [\ion{S}{3}] line emission reaching a couple of thousand Jy at the location of position 2 where the 20\,$\mu$m flux is highest. It should be stated that all of these \textit{ISO} spectra were sampled within a variable aperture size from 14$\arcsec\times$27$\arcsec$ at wavelengths around 20\,$\mu$m to 20$\arcsec\times$33$\arcsec$ around 37\,$\mu$m. Given such large apertures, it is likely that the line strengths vary quite considerably within subregions of the aperture and could be much higher than seen in the spectra presented. 

The bottom of Figure \ref{fig:app} also shows a 15--42\,$\mu$m spectrum of position 2 with the 20 and 37\,$\mu$m FORCAST filter profiles overlaid. The transmission profiles of the filters are normalized with their peak transmissions at 100\%, and take into account all transmission elements below the atmosphere (i.e. telescope and instrument optics, and detector quantum efficiency as a function of wavelength). It can be seen from this figure that the only non-continuum emission source in the 20\,$\mu$m passband is the [\ion{S}{3}] line at 18.71\,$\mu$m. 

The FORCAST 37\,$\mu$m filter is actually quite broad and there was some concern as to why this filter was not affected as well, since there is an even brighter [\ion{S}{3}] line at 33.48\,$\mu$m. Figure \ref{fig:app} shows that the FORCAST 37\,$\mu$m filter has only a few percent transmission at 33.48\,$\mu$m, and the other lines present in its passband ([\ion{Ne}{3}] at 36.01\,$\mu$m and [\ion{Si}{2}] at 34.81\,$\mu$m) are rather weak. Therefore, the 37\,$\mu$m filter can be considered to be dominated by dust continuum emission. 

This enhanced flux measured in regions of ionized emission may not be unique to FORCAST observations at 20\,$\mu$m. Most mid-infrared instruments have a filter centered near 20\,$\mu$m, and so caution should be taken when using data from similar filters. For instance, a comparison of surface brightnesses derived from the 22\,$\mu$m (Band E) data from the the \textit{Midcourse Space Experiment (MSX)} to our FORCAST 20\,$\mu$m surface brightnesses for other ionized sources we are studying show similar values. This is likely because this \textit{MSX} filter has a bandpass ($\sim$18 to $\sim$25\,$\mu$m) that also encompasses the [\ion{S}{3}] line at 18.71\,$\mu$m. However, the \textit{Wide-Field Infrared Survey Explorer (WISE)} 22\,$\mu$m (Band 4) and \textit{Spitzer}-MIPS 24\,$\mu$m filters do not have passbands that would be affected by this line.

\end{document}